\documentclass{article}
\usepackage[utf8]{inputenc}
\usepackage{amsmath,amssymb}
\usepackage{bm, bbm, blkarray, adjustbox}
\usepackage{amsfonts}
\usepackage{enumitem}
\usepackage[margin=1in]{geometry}
\usepackage{natbib}
\usepackage{hyperref}

\usepackage{graphicx}
\usepackage{soul}
\usepackage{setspace}
\usepackage{listings}
\usepackage{color,soul}
\usepackage{authblk}
\usepackage{multirow}
\usepackage{booktabs}
\usepackage{pdflscape}
\usepackage{xcolor}
\usepackage{algorithm,algorithmicx,algpseudocode}

\renewcommand{\d}[1]{\ensuremath{\operatorname{d}\!{#1}}}

\title{Generalized promotion time cure model:\\ A new modeling framework to identify cell-type-specific genes and improve survival prognosis}
\author[1]{Zhi Zhao}
\author[1]{Fatih K\i z\i laslan}
\author[2]{Shixiong Wang}
\author[1,3]{Manuela Zucknick}

\affil[1]{Oslo Centre for Biostatistics and Epidemiology (OCBE), University of Oslo, Norway}
\affil[2]{Department of Clinical Molecular Biology(EpiGen), Akershus University Hospital, Norway}
\affil[3]{Oslo Centre for Biostatistics and Epidemiology (OCBE), Oslo University Hospital, Norway}

\date{}

\begin{document}

\maketitle

\begin{abstract}
Single-cell technologies provide an unprecedented opportunity for dissecting the interplay between the cancer cells and the associated tumor microenvironment, and the produced high-dimensional omics data should also augment existing survival modeling approaches for identifying tumor cell type-specific genes predictive of cancer patient survival. 
However, there is no statistical model to integrate multiscale data including individual-level survival data, multicellular-level cell composition data and cellular-level single-cell omics covariates. 
We propose a class of Bayesian generalized promotion time cure models (GPTCMs) for the multiscale data integration to identify cell-type-specific genes and improve cancer prognosis. 
We demonstrate with simulations in both low- and high-dimensional settings that the proposed Bayesian GPTCMs are able to identify cell-type-associated covariates and improve survival prediction. 
\end{abstract}


\section{Introduction}

A fundamental aspect of cancer research is to identify and study the factors driving cancer progression. However, despite the identification of hundreds of potential molecular prognostic and treatment-predictive biomarkers for cancer through whole-genome or whole-transcriptome studies in the past two decades, emergent drug resistance often causes the therapies to fail even after initial successful rounds of (partial) response, which is largely due to a key property of cancer, which is intratumor heterogeneity 
\citep{DagogoJack2017,McGranahan2017,Zhang2022}. 
The heterogeneity between cancerous cells, i.e., the fact that different cancer cells within the same tumor might have originated from different subclones with different subclonal driver alterations, or for example in the case of B-cell lymphoma represent the heterogeneity of the malignant cells (B lymphocytes) as well as their tumor microenvironment, with cell-type-specific omic features, may play a substantial role in the cancer progression or prognosis. 

Single-cell technologies provide an unprecedented opportunity for dissecting the interplay between the cancer cells and the associated tumor microenvironment (TME), and the produced high-dimensional omics data should also augment existing survival modeling approaches for identifying tumor cell type-specific genes predictive of cancer patient survival. 
Current literature and contemporary insights from cancer biology reveal that deciphering the composition of the TME is critical for understanding tumorigenesis, prognosis and responsiveness to immunotherapies \citep{Cai2024}, but current computational approaches often map the findings from single-cell omics features to large-scale bulk sequencing omics features and then jointly analyzing the bulk omics and survival data, rather than directly modeling single-cell omics and survival data \citep{Zhao2024survomics}. 
Although there are multiple survival models \citep{Yakovlev1996,Chen1999,Cooner2007,Kim2011,Gomez2023} motivated by the growth of clonogenic tumor cells, they cannot jointly model individual-level survival data and (multi-)cellular-level omics data. 

Our previous work \cite{Zhao2024gptcm} proposed a new survival model, 
generalized promotion time cure model (GPTCM), for multiscale data integration including individual-level survival data, multicellular-level cell composition data and cellular-level single-cell omics covariates. 
In this article, we develop a class of Bayesian versions of GPTCM to identify cell-type-specific genes and improve survival prognosis. 
We also present complete statistical inference for the proposed Bayesian GPTCMs, and investigate the model performance through systematic simulations in both low- and high-dimensional settings. 
An R package {\bf GPTCM} \citep{Zhao2025cran} is available from the Comprehensive R Archive Network at \url{https://CRAN.R-project.org/package=GPTCM}.

The rest of the article is organized as follows. 
In Section \ref{sec:methods}, we propose six Bayesian versions of GPTCM and present Bayesian inference for the estimation of the proposed models. 
In Section \ref{sec:simulations}, we compare the performances of the proposed Bayesian GPTCMs and classical survival models through simulation in low- and high-dimensional settings. 
In Section \ref{sec:discussion}, we conclude the article with a discussion.

\section{Generalized promotion time cure model}\label{sec:methods}

\subsection{Framework of GPTCM}

The generalized promotion time cure model (GPTCM) \citep{Zhao2024gptcm} was developed based on the classical promotion time cure model (PTCM) \citep{Yakovlev1996}. 
Instead of assuming all clonogenic tumor cells homogeneous in PTCM, the generalized version, GPTCM, assumes that the clonogenic tumor cells are a composition of multiple cancer cell subtypes in order to capture intra-tumor heterogeneity for better modeling tumor evolution and for better predicting cancer patient survival. 
Suppose that a patient after an initial treatment has the total count of cancer cells $N = \sum_{l=1}^L N_l$, $L \ge 2$, where $N_l$ is the number of cells in the $l$-th cancer cell subtype, and suppose the total count $N\sim \mathcal Poisson(\theta)$ with $\theta>0$. 
Given the total count $N$, the counts of individual cancer cell subtypes follow a multinomial distribution, i.e., $(N_1,...,N_L)\sim \mathcal Mult(\mathtt p_1,...,\mathtt p_L)$, where the sum of the $L$ probabilities is $\sum_{l=1}^L\mathtt p_l=1$.  
To generalize the PTCM, suppose the $l$-th cell type has multivariate random times for its $N_l$ cancer cells propagating into a newly detectable tumor:
$$
\bm W_l = \left(Z_{\sum_{j=1}^{l-1}N_{j-1}+1}, ..., Z_{\sum_{j=1}^{l-1}N_{j-1}+N_l}\right),
$$ 
where $Z_{\bullet}$ is the random time for a clonogenic cell to produce a detectable tumor mass, $l\in \{1,...,L\}$, $N_0=0$. 
For the $N_l$ homogeneous cells in the $l$-th cell type, we assume a cell-type-specific promotion time distribution $F_l(t)=1-S_l(t)$, $l\in \{1,...,L\}$. 
The time to tumor recurrence can be defined as 
$T=\min \{\min\{\bm W_1\},...,\min\{\bm W_L\}\}$, i.e., the time when the first clonogenic cell in one cell type becomes activated. 
Then \cite{Zhao2024gptcm} derived the population survival function as
\begin{align*} 
  S_{pop}(t) &= 
  \mathbb P(\text{no cancer by time }t) \\
  &= \mathbb P(N=0) + \mathbb P\left(\min\{\min\{\bm W_1\},...,\min\{\bm W_L\}\} >t, N>0\right) \\
  &= e^{-\theta\left\{1 - \sum_{l=1}^L\mathtt p_lS_l(t) \right\}}
  . \tag{1}\label{eq:gptcm}
\end{align*}
If the promotion time distributions of different cell types are the same, i.e., $S(t)=S_l(t)$, $\forall l\in \{1,...,L\}$, the population survival function (\ref{eq:gptcm}) is degenerated into PTCM \citep{Yakovlev1996}, i.e., $S_{pop}(t) = e^{-\theta\left\{1 - S(t) \right\}}$. 
Note that the population survival function will not become zero when the time tends to infinity, i.e., some patients will never encounter the event of interest. 
Patients who will never encounter the event of interest belong to the cure population (with a fraction $e^{-\theta}$), and patients who will potentially encounter the event of interest always after a finite time point belong to the noncure population. 

In both PTCM and GPTCM, covariates can be introduced into the Poisson rate parameter $\theta$ through a log-linear submodel:
\begin{equation*} \tag{2}\label{eq:gptcm-cure}
  \log\theta = \xi_0+ \mathbf X_0\bm\xi,
\end{equation*}
where $\mathbf X_0\in \mathbb R^{n\times d}$ is a data matrix of $d$ mandatory variables from $n$ subjects, and coefficients $\bm\xi=(\xi_1,...,\xi_d)^\top$. 
As indicated in \cite{Zhao2024gptcm}, benefiting from the mixture part $\sum_{l=1}^L\mathtt p_lS_l(t)$ in GPTCM (\ref{eq:gptcm}), cell-type-specific covariates (e.g. genetic variables from each cancer cell subtype) $\mathbf X_l\in\mathbb R^{n\times p}$ can be introduced into the survival function $S_l(t)$. 
For example, we can model a Weibull's mean parameter $\bm\mu_l\in\mathbb R^n$ by cell-type-specific covariates $\mathbf X_l$: 
\begin{align*}
S_l(t) &= \exp\{ -(t / \bm\lambda_l)^\kappa \}, \\
\bm\lambda_l &= \frac{\bm\mu_l} {\Gamma(1+1/\kappa)},\\
\log\bm\mu_l &= \beta_{0l} + \mathbf X_l\bm\beta_l,
\end{align*}
where $\kappa \in \mathbb R_+$ is the Weibull's shape parameter and $\Gamma(\cdot)$ is the gamma function. 

We will propose six Bayesian versions of GPTCM as shown in Table \ref{tab:gptcm} suited for modeling low- and high-dimensional covariates, and present full Bayesian inference for the six models. 
For the simplicity of notations, we assume the same number of covariates from different cell types, i.e., $\mathbf X_1,...,\mathbf X_L$ have the same number of columns. 
However, the proposed models can deal with different numbers of covariates from different cell types. 

\begin{landscape}
  \thispagestyle{empty}
\begin{table}[!htbp]
\centering
\caption{Bayesian GPTCMs\medskip\label{tab:gptcm} }
\medskip
{\scriptsize %
  \makebox[\textwidth][c]{
\begin{tabular}{l c | c | c | c | c | c}  
\hline\hline
  &
  GPTCM-noBVS1 &
  GPTCM-noBVS2 & 
  GPTCM-Ber1 & 
  GPTCM-Ber2 & 
  GPTCM-MRF1 & 
  GPTCM-MRF2  
\smallskip
\\
\hline
Survival function & \multicolumn{6}{l}{}  \\
& \multicolumn{6}{c}{
  $\begin{aligned}
    S_{pop}(t) &= e^{ -\theta\{1 - \sum_{l=1}^L \mathtt p_l S_l(t)\} } \\
    S_l(t) &= e^{-\left(t/\lambda_l\right)^\kappa}, \ l=1,...,L \\ 
    \lambda_l &= \frac{\bm\mu_l}{\Gamma(1+1/\kappa)}\\
    \kappa &\sim \mathcal Gamma(a_\kappa, b_\kappa)
  \end{aligned}$
} \\
\cmidrule{2-7}
Cure fraction & \multicolumn{6}{l}{}  \\
& \multicolumn{6}{c}{
  $\begin{aligned}
    \log\theta &= \xi_0 + \mathbf X_0\bm\xi, \text{ where } \bm\xi=(\xi_1,...,\xi_d)^\top\\
    \xi_0 &\sim \mathcal N(0, v_0^2), \ v_0^2 \sim \mathcal{IG}amma(a_{v_0}, b_{v_0})\\ 
    \xi_k &\sim \mathcal N(0, v^2), \ v^2 \sim \mathcal{IG}amma(a_v, b_v)
  \end{aligned}$
} \\
\cmidrule{2-7}
Noncure fraction & \multicolumn{6}{l}{}  \\
& \multicolumn{6}{c}{
  $\begin{aligned}
  \log\bm\mu_l &= \beta_{0l} + \mathbf X_l\bm\beta_l, \text{ where } \bm\beta_l=(\beta_{1l},...,\beta_{pl})^\top\\
  \beta_{0l} &\sim \mathcal N(0, \tau_0^2)
  \end{aligned}$
} 
\bigskip\\
&
\multicolumn{2}{c}{
  $\begin{aligned}
&\text{\hspace{-8mm}Normal prior:} \\
    \beta_{jl} &\sim \mathcal N(0, \tau_l^2), \ \ j=1,...,p,\ l=1,...,L\\
    \tau_l^2 &\sim \mathcal{IG}amma(a_\tau, b_\tau)
    \\ \\ \\ \\ 
  \end{aligned}$
} &
\multicolumn{2}{|c|}{
$\begin{aligned}
&\text{\hspace{-23mm}Spike-and-slab prior:} \\
  \beta_{jl} | \gamma_{jl}, \tau_l^2 &\sim \gamma_{jl}\mathcal N(0, \tau_l^2) + (1-\gamma_{jl})\delta_0(\beta_{jl})\\
&\text{\hspace{-23mm}Bernoulli-beta:} \\
  \gamma_{jl}|\pi_{jl} &\sim \mathcal Bernoulli (\pi_{jl}) \\
  \pi_{jl} &\sim \mathcal Beta(a_\pi, b_\pi)\\
  \tau_l^2 &\sim \mathcal{IG}amma(a_\tau, b_\tau)
\end{aligned}$
}
& 
\multicolumn{2}{c}{
$\begin{aligned}
&\text{\hspace{-25mm}Spike-and-slab prior:} \\
  \beta_{jl} | \gamma_{jl}, \tau_l^2 &\sim \gamma_{jl}\mathcal N(0, \tau_l^2) + (1-\gamma_{jl})\delta_0(\beta_{jl}) \\
&\text{\hspace{-25mm}MRF prior:} \\
 f(\bm\gamma|a,b,G) &\propto \exp\{a\mathbbm{1}^\top\bm\gamma+ b\bm\gamma^\top G \bm\gamma\} \\ 
 \bm\gamma &= (\gamma_{11},\gamma_{12},...,\gamma_{pL})^\top \\
  \tau_l^2 &\sim \mathcal{IG}amma(a_\tau, b_\tau) 
  \\
\end{aligned}$
} \\
\cmidrule{2-7}
Measurement error & \multicolumn{6}{l}{}  \\
& 
\multicolumn{1}{c}{
$\begin{aligned}
  \tilde{\mathbf p} &= \mathbf p + \text{No error} \\
  \\
&\text{\hspace{-13mm}Observed proportions:} \\
  \tilde{\mathbf p} &= [\tilde{\mathtt p}_1,...,\tilde{\mathtt p}_L] \\
  \\
&\text{\hspace{-13mm}Underlying proportions:} \\
  \mathbf p &= [\mathtt p_1,...,\mathtt p_L] \\
\end{aligned}
$
} 
& 
\multicolumn{1}{|c|}{
$\begin{aligned}
  \tilde{\mathbf p} &= \mathbf p + \text{Error}\\
  \tilde{\mathbf p} &\sim \mathcal Dir(\bm\alpha), \ \ 
  \mathbb E[\tilde{\mathbf p}] = \mathbf p\\
  \bm\alpha &= (\alpha_1,...,\alpha_L) \\
  \log\alpha_l &= \zeta_{0l} + \mathbf X_l\bm\zeta_l \\
  \zeta_{0l} | w_0^2 &\sim \mathcal N(0, w_0^2) \\
  \\
&\text{\hspace{-9mm}Normal prior:} \\
  \bm\zeta_l &= (\zeta_{1l},...,\zeta_{pl})^\top \\
  \zeta_{jl} | w_l^2 &\sim \mathcal N(0, w_l^2) \\
  w_l^2 &\sim \mathcal{IG}amma(a_w, b_w)
  \\
\end{aligned}$
} 
&
\multicolumn{1}{c}{
$
  \tilde{\mathbf p} = \mathbf p + \text{No error}
$
} 
&
\multicolumn{1}{|c|}{
$\begin{aligned}
  \tilde{\mathbf p} &= \mathbf p + \text{Error}\\
  \tilde{\mathbf p} &\sim \mathcal Dir(\bm\alpha), \ \ 
  \mathbb E[\tilde{\mathbf p}] = \mathbf p\\
  \log\alpha_l &= \zeta_{0l} + \mathbf X_l\bm\zeta_l \\
  \zeta_{0l} | w_0^2 &\sim \mathcal N(0, w_0^2) \\
  \\
&\text{\hspace{-13mm}Spike-and-slab prior:} \\
  \zeta_{jl} | \eta_{jl}, w_l^2 &\sim \eta_{jl}\mathcal N(0, w_l^2) + (1-\eta_{jl})\delta_0(\zeta_{jl}) \\
&\text{\hspace{-13mm}Bernoulli-beta:} \\
  \eta_{jl}|\rho_{jl} &\sim \mathcal Bernoulli (\rho_{jl}) \\
  \rho_{jl} &\sim \mathcal Beta(a_\rho, b_\rho)\\
  w_l^2 &\sim \mathcal{IG}amma(a_w, b_w)
\end{aligned}$
} 
&
\multicolumn{1}{c}{
$
  \tilde{\mathbf p} = \mathbf p + \text{No error}
$
} 
&
\multicolumn{1}{|c}{
$\begin{aligned}
  \tilde{\mathbf p} &= \mathbf p + \text{Error}\\
  \tilde{\mathbf p} &\sim \mathcal Dir(\bm\alpha), \ \ 
  \mathbb E[\tilde{\mathbf p}] = \mathbf p\\
  \log\alpha_l &= \zeta_{0l} + \mathbf X_l\bm\zeta_l \\
  \zeta_{0l} | w_0^2 &\sim \mathcal N(0, w_0^2) \\
  \\
&\text{\hspace{-18mm}Spike-and-slab prior:} \\
  \zeta_{jl} | \eta_{jl}, w_l^2 &\sim \eta_{jl}\mathcal N(0, w_l^2) + (1-\eta_{jl})\delta_0(\zeta_{jl}) \\
&\text{\hspace{-18mm}MRF prior:} \\
  f(\bm{\eta}|a^\star,b^\star,G^\star) &\propto \exp\{a^\star\mathbbm{1}^\top\bm{\eta}+ b^\star\bm{\eta}^\top G^\star \bm{\eta}\} \\
  w_l^2 &\sim \mathcal{IG}amma(a_w, b_w) \\
  \\
\end{aligned}$
} 
\smallskip\\
\hline\hline
\end{tabular}
}%
}
\end{table}

\end{landscape}

\subsection{Bayesian GPTCMs}
\medskip
{\bf GPTCM-noBVS1} 
\smallskip\\
A simple version of GPTCM with covariates (denoted as GPTCM-noBVS1 in Table \ref{tab:gptcm}) is to introduce $d$ mandatory variables $\mathbf X_0\in\mathbb R^{n\times d}$ into the cure fraction through the log-linear submodel (\ref{eq:gptcm-cure}), and introduce cell-type-specific covariates into the Weibull's mean parameter $\bm\mu_l$ through another log-linear submodel:
\begin{equation}\tag{3}\label{eq:weibull_mu}
 \log\bm\mu_l = \beta_{0l} + \mathbf X_l\bm\beta_l,\ \ l=1,...,L,
\end{equation}
where $\mathbf X_l \in\mathbb R^{n\times p}$ includes $p$ cell-type-specific covariates, and coefficients $\bm\beta_l=(\beta_{1l},...,\beta_{pl})^\top$. 
By convention, for the log-linear submodel (\ref{eq:gptcm-cure}), a normal prior is used for the intercept $\xi_0\sim\mathcal N(0, v_0^2)$, and another normal prior for each variable's effect, i.e., $\xi_k\sim\mathcal N(0, v^2)$, $k=1,...,d$. 
Similarly, for the log-linear submodel (\ref{eq:weibull_mu}), normal priors are used for $\beta_{0l}$ and $\beta_{jl}$:
$$
\beta_{0l} \sim \mathcal N(0, \tau_0^2), \ \ 
\beta_{jl} \sim \mathcal N(0, \tau_l^2), \ \ 
l = 1,...,L, \ \ j = 1,...,p.
$$
Here, the cell-type-specific prior variances $\tau_l^2$ allow different scales of cell-type-specific effects. 
Independent inverse-gamma priors can be used for all variances in the normal priors. 
This Bayesian version of GPTCM can be suited to handle low-dimensional covariates without the need of Bayesian variable selection (BVS). 
\medskip\\
{\bf GPTCM-noBVS2} 
\smallskip\\
The GPTCM (\ref{eq:gptcm}) assumes that after an initial treatment the numbers of multiple cancer cell subtypes follow a multinomial distribution assuming that the cell type proportions $\mathbf p = [\mathtt p_1,...,\mathtt p_L]$ are known. 
However, the true cell type proportions may not be measured accurately in real applications. 
We can assume that the measured proportions of multiple cell types $\tilde{\mathbf p}= [\tilde{\mathtt p}_1,...,\tilde{\mathtt p}_L]$ follow a Dirichlet distribution, $\mathcal Dir(\bm\alpha)$, with concentration parameters $\bm\alpha=(\alpha_1,...\alpha_l,...,\alpha_L)$, $\alpha_l>0$, and with expectations as the underlying probabilities $\mathbf p$. 
Then we can use a measurement error model for the measured proportions data $\tilde{\mathbf p}$:
\begin{align*}
    \tilde{\mathbf p} \sim \mathcal Dir(\bm\alpha), \ \
  \mathtt p_l = \mathbb E[\tilde{\mathtt p}_l] = \frac{\alpha_l}{\sum_{l'=1}^L\alpha_{l'}},
\end{align*}
An additional advantage of this model is that we can introduce cell-type-specific covariates to model the cell type proportions; they can be introduced into the Dirichlet distribution via its concentration parameters similar to Dirichlet regression with the common parametrization \citep{Maier2014}: 
$$
\log\alpha_l = \zeta_{0l} + \mathbf X_l\bm\zeta_l,\ \ l = 1,...,L.
$$
The idea of using the measurement error submodel is similar to the joint models \citep{Faucett1996}. 
Similar to GPTCM-noBVS1, this version of GPTCM (denoted as GPTCM-noBVS2 shown in Table \ref{tab:gptcm}) can be suited to handle low-dimensional covariates without BVS. 
\medskip\\
{\bf GPTCM-Ber1} 
\smallskip\\
In order to be able to analyze data with high-dimensional covariate space, particularly for identifying cell-type specific genes relevant in cancer research, we can employ Bayesian variable selection, e.g. through independent spike-and-slab priors for the coefficients $\beta_{jl}$ linked to the mean parameters of cell-type-specific Weibull distributions in (\ref{eq:weibull_mu}): 
\begin{align*} \tag{4}\label{eq:spikeslab}
  \beta_{jl} | \gamma_{jl}, \tau_l^2 &\sim \gamma_{jl}\mathcal N(0, \tau_{l}^2) + (1-\gamma_{jl})\delta_0(\beta_{jl}), \\
  \gamma_{jl} &\sim \mathcal Bernoulli (\pi_{jl}),
\end{align*}
where $\gamma_{jl}$ ($l=1,\cdots,L; j=1,\cdots,p$) is a latent variable (with a Bernoulli hyperprior) for variable selection indicating $\beta_{jl}\neq 0$ if $\gamma_{jl}=1$ and $\beta_{jl}= 0$ if $\gamma_{jl}=0$, $\tau_{l}^2$ is a shrinkage parameter with hyperprior $\tau_{l}^2\sim \mathcal{IG}(a_\tau,b_\tau)$, and $\delta_0(\cdot)$ is the Dirac delta function. 
We specify a beta hyperprior on the Bernoulli's probability $\pi_{jl}\sim \mathcal Beta(a_\pi, b_\pi)$. 
\medskip\\
{\bf GPTCM-Ber2} 
\smallskip\\
In addition to the Bayesian variable selection for the coefficients $\beta_{jl}$ as in GPTCM-Ber1, we can also implement BVS in the same way for covariates linked to the cell-type-specific proportions in the measurement error submodel: 
\begin{align*}
  \log\alpha_l &= \zeta_{0l} + \mathbf X_l\bm\zeta_l, \\
  \zeta_{jl} | \eta_{jl}, w_l^2 &\sim \eta_{jl}\mathcal N(0, w_l^2) + (1-\eta_{jl})\delta_0(\zeta_{jl}), \\
  \eta_{jl}|\rho_{jl} &\sim \mathcal Bernoulli (\rho_{jl}), \\
  \rho_{jl} &\sim \mathcal Beta(a_\rho, b_\rho),
\end{align*}
where $\eta_{jl}$'s are variable selection indicators for covariates linked to the Dirichlet's concentration parameters, and the Bernoulli-beta prior is also used for $\eta_{jl},\ l = 1,...,L,\ j = 1,...,p$. 
This model is denoted as GPTCM-Ber2 shown in Table \ref{tab:gptcm}. 
\medskip\\
{\bf GPTCM-MRF1} 
\smallskip\\
As an alternative to the purely data-driven variable selection via the spike-and-slab priors in (\ref{eq:spikeslab}), we can make use of prior biological knowledge, for example of intercellular networks/pathways by employing graph-structured priors for variable selection. 
A Markov random field prior (MRF) \citep{Li2010,Stingo2011,Zhao2024mrf} can be used for the variable selection indicators of the effects $\bm\beta_l$ in (\ref{eq:weibull_mu}) for variable selection and simultaneously address known relationships between 
cell-type-specific covariates. 
The MRF prior is defined for the vector of all variable selection indicators, i.e., $\bm\gamma = \text{vec}[(\bm\gamma_1,...,\bm\gamma_l,...,\bm\gamma_L)]=(\gamma_{11},\gamma_{21},...,\gamma_{pL})^\top$, $\bm\gamma_l=(\gamma_{1l},...,\gamma_{pl})^\top$, $l=1,...,L$: 
\begin{equation*} \tag{5}\label{eq:mrf}
 f(\bm\gamma|a,b,G) \propto \exp\{a\mathbbm{1}^\top\bm\gamma+ b\bm\gamma^\top G \bm\gamma\},
\end{equation*}
where $a<0$ controls overall model sparsity, $b\ge 0$ determines the strength of the structural relationships between variables, and $G$ is a $pL \times pL$ matrix representing a (weighted) graph corresponding to the known structural relationships between variables. 
If $b=0$, the MRF prior is degenerated to independent Bernoulli priors for individual variable selection indicators. 
The full model is denoted as GPTCM-MRF1 shown in Table \ref{tab:gptcm}.
\medskip\\
{\bf GPTCM-MRF2} 
\smallskip\\
Similarly to the extension of GPTCM-Ber1 to GPTCM-Ber2, we can again also employ the graph-structured priors for variable selection of covariates linked to cell-type-specific proportions in the measurement error submodel and thus extend from GPTCM-MRF1 to GPTCM-MRF2: 
\begin{equation*}
 f(\bm\eta|a^\star,b^\star,G^\star) \propto \exp\{a^\star\mathbbm{1}^\top\bm\eta+ b^\star\bm\eta^\top G^\star \bm\eta\},
\end{equation*}
where the vector of all variable selection indicators $\bm\eta = \text{vec}[(\bm\eta_1,...,\bm\eta_l...,\bm\eta_L)]=(\eta_{11},\eta_{21},...,\eta_{pL})^\top$, $\bm\eta_l=(\eta_{1l},...,\eta_{pl})^\top$, parameters $a^\star$ and $b^\star$ control the model sparsity and strength of prior graph knowledge $G^\star$, respectively. 
Note that the graph $G^\star$ here can be the same as the graph $G$ in (\ref{eq:mrf}), if we do not have any prior biological knowledge that can help us to distinguish the relationships between the cell-type-specific survival related effects and the cell-type-specific proportions related effects. 
We can also combine the two graphs into diag$\{G,G^\star\}$ and add additional nonzero weights in its off-diagonal blocks to encourage joint selection between variables linked to cell-type-specific survival and variables linked to cell-type-specific proportions. 
The full model is denoted as GPTCM-MRF2 shown in Table \ref{tab:gptcm}.

\subsection{Computation}\label{sec:computation}

We consider a survival modeling framework for right-censored time-to-event data. 
Let $T_i$ and $C_i$, $i=1,\cdots,n$, be the times to tumor progression and censoring for the $i$-th subject, respectively. 
The observed time $t_i = \min\{T_i, C_i\}$ and the censoring indicator $\delta_i = \mathbbm 1 \{T_i \le C_i\}$. 
We consider commonly used assumptions: (i) independence between $T_i$ and $C_i$, (ii) independence between $(T_i, C_i)$ and $(T_j,C_j)$ for $1 \le i \ne j \le n$, (iii) non-informative censoring, and cure rate $\mathbb P(N=0)=e^{-\theta}$, $\theta>0$. 
Let $\bm{\mathcal X}_i=\{\mathbf X_{i0}, \mathbf X_{i1},...,\mathbf X_{il},...,\mathbf X_{iL}\}$ be time-independent clinical covariates and cell-type-specific covariates. 
The joint distribution of observed time to an event and observed proportions can be decomposed into the product of the distribution of observed time to an event conditional on the observed proportions and the distribution of the observed proportions, i.e.,
$$
f(t_i,\tilde{\mathbf p}_i| \mathbf p_i, \bm{\mathcal X}_i) 
\propto 
f(t_i|\tilde{\mathbf p}_i, \mathbf p_i, \bm{\mathcal X}_i)f(\tilde{\mathbf p}_i| \mathbf p_i, \bm{\mathcal X}_i)
=
f(t_i|\mathbf p_i, \bm{\mathcal X}_i)f(\tilde{\mathbf p}_i| \bm{\mathcal X}_i).
$$
Note that in $f(t_i|\mathbf p_i, \bm{\mathcal X}_i)$, $\mathbf p_i=\mathbb E[\tilde{\mathbf p}_i]$ will be estimated by the Dirichlet regression fitted responses. 
The joint likelihood function for the entire population is
\begin{align*} \tag{6}\label{eq:likelihood}
  \mathcal L(\bm\vartheta | \mathcal D) 
  \propto 
  \prod_{i=1}^{n} f_{pop}(t_i|\bm{\mathcal X}_i,\mathbf p_i)^{\delta_i} 
  S_{pop}(t_i|\bm{\mathcal X}_i,\mathbf p_i)^{{1-\delta_i}} f(\tilde{\mathbf p}_i | \bm{\mathcal X}_i), 
\end{align*}
where $\bm\vartheta$ consists of all relevant parameters, $\mathcal D=\{t_i, \delta_i, \bm{\mathcal X}_i, \tilde{\mathbf p}_i\}_{i=1}^n$ is the observed data, and 
$
f_{pop}(t_i|\bm{\mathcal X}_i,\mathbf p_i) 
= 
-\frac{\d{}}{\d t} S_{pop}(t_i|\bm{\mathcal X}_i,\mathbf p_i)
$ 
is the probability density function corresponding to the population survival function. 
Note that $S_{pop}(t)$ is not a proper survival function, so $f_{pop}(t_i|\bm{\mathcal X}_i)$ is not a proper probability density function. 

To proceed with posterior computation for GPTCM-MRF2 as an example, we assume mutual independence among the prior of the Weibull's shape parameter and the priors of the coefficients in cure and noncure fractions. 
The joint posterior distribution is composed by the likelihood (\ref{eq:likelihood}) and the joint distribution of hyperparameters, i.e.,
\begin{align*}
  f(\bm\vartheta |\mathcal D) 
  &\propto \mathcal L(\bm\vartheta | \mathcal D) f(\bm\vartheta) \\
  &= \mathcal L(\bm\vartheta | \mathcal D) f(\bm\xi, \xi_0, v^2, v_0^2, \kappa, \bm\beta, \bm\beta_0, \bm\tau^2, \tau_0^2, \bm\gamma, G, \bm\zeta, \bm\zeta_0, \bm w^2, w_0^2, \bm\eta, G^\star) \\
  &= \mathcal L(\bm\vartheta | \mathcal D) 
  \prod_{k=1}^d \{f(\xi_k|v^2)\}f(v^2) f(\xi_0|v_0^2)f(v_0^2) f(\kappa)
  \prod_{l=1}^L \{
    f(\bm\beta_l | \bm\gamma_l, \bm\tau_l) f(\bm\tau_l) f(\beta_{0l}|\tau_0^2)f(\tau_0^2)
  \} 
    f(\bm\gamma | G) \\
  & \ \ \ \times 
  \prod_{l=1}^L \{
    f(\bm\zeta_l | \bm\eta_l, \bm w) f(\bm w) f(\zeta_{0l}|w_0^2)f(w_0^2)
  \} 
    f(\bm\eta | G^\star) \\
  &= \prod_{i=1}^{n} f_{pop}(t_i|\bm{\mathcal X}_i,\mathbf p_i)^{\delta_i} 
  S_{pop}(t_i|\bm{\mathcal X}_i,\mathbf p_i)^{{1-\delta_i}} 
  f(\tilde{\mathbf p}_i | \bm{\mathcal X}_i) 
    \prod_{k=1}^d \{f(\xi_k|v^2)\}f(v^2) f(\xi_0|v_0^2)f(v_0^2) f(\kappa) \\
  & \ \ \ \times  
  \prod_{l=1}^L \left\{ \prod_{j=1}^p
    \{ f(\beta_{jl} | \gamma_{jl}, \tau_{l}^2) f(\tau_{l}^2) 
    f(\beta_{0l}|\tau_0^2) f(\tau_0^2) 
    \} \right\} f(\bm\gamma | G) \\
  & \ \ \ \times  
  \prod_{l=1}^L \left\{ \prod_{j=1}^p
    \{ f(\zeta_{jl} | \eta_{jl}, w_l^2) f(w_l^2) 
    f(\zeta_{0l}|w_0^2) f(w_0^2) 
    \} \right\} f(\bm\eta | G^\star).
\end{align*}
All relevant full conditionals can be found in Supplementary material \ref{secSuppl:posterior}. 
For simple full conditionals with beta and inverse-gamma distributions, we use Gibbs sampling. 
Since the full conditionals of $\bm\beta=\text{vec}[(\bm\beta_1,...,\bm\beta_L)]$ and $\bm\zeta=\text{vec}[(\bm\zeta_1,...,\bm\zeta_L)]$ are log-concave and can be for high-dimensional parameters, we update them by using the adaptive rejection Metropolis sampling (ARMS) \citep{Gilks1995} conditional on corresponding nonzero variable selection indicators. 
For computational efficiency, at every Markov chain Monte Carlo (MCMC) iteration we update the pair of coefficients and corresponding variable selection indicators, i.e., $(\bm\beta_l, \bm\gamma_l)$ or $(\bm\zeta_l,\bm\eta_l)$, for one randomly selected $l$-th cell type. 
The pair $(\bm\beta_l, \bm\gamma_l)$ (or $(\bm\zeta_l,\bm\eta_l)$) is jointly updated via the Metropolis-Hastings (M-H) sampling, with proposal variable selection indicators using a multi-armed bandits technique \citep{Thompson1933,Sutton1998}, and with the proposal coefficients based on the ARMS conditional on the proposal variable selection indicators, see Supplementary material \ref{secSuppl:posterior} for details. 
For other full conditionals with complex densities, we use slice sampling \citep{Neal2003}. 
Algorithm \ref{alg1} shows the MCMC for fitting GPTCM-MRF2. 
The algorithm is also used for other Bayesian versions of GPTCM with slight changes based on different priors. 
\begin{algorithm}[H]
{\small
  \caption{\label{alg1}: MCMC for GPTCM-MRF2}
  \begin{algorithmic}[1]
   \State Set hyperparameters 
   $a_v$, $b_v$, $a_{v_0}$, $b_{v_0}$, 
   $a_\tau$, $b_\tau$, $a_{\tau_0}$, $b_{\tau_0}$, 
   $a_w$, $b_w$, $a_{w_0}$, $b_{w_0}$, 
   $a_\kappa$, $b_\kappa$, 
   $G$, $G^\star$, $a$, $b$, $a^\star$, $b^\star$
   \State Initialize coefficients $\xi_0$, $\bm\xi$, $\bm\beta_l$ and $\bm\gamma_l$ ($l=1,...,L$)
   \State Set the number of iterations $M$
   \For{\texttt{$i = 1,\cdots,M$ }}
      \State Draw $v_0^2$ and $v^2$ via Gibbs sampling
      \State Draw $\xi_0$ and $\xi_k$ ($k=1,...,d$) via slice sampling
      \State Update cure fraction rate $\bm\theta = \exp\{\xi_0+\mathbf X_0\bm\xi\}$
      \State Randomly select one $l$ from $\{1,...,L\}$, and 
      draw $(\bm\zeta_l,\bm\eta_l)$ via M-H sampling
      \State Draw all $\bm\zeta_l$ via ARMS, $\forall l$ 
      \State Update Dirichlet mean proportions $\mathtt p_l = \frac{\alpha_l}{\sum_{l'=1}^L\alpha_{l'}}$, where $\alpha_l = \exp\{\zeta_{0l} + \mathbf X_l\bm\zeta_l\}$, $\forall l$
      \State Draw Weibull's shape parameter $\kappa$ via slice sampling, and update $\bm\lambda_l=\frac{\bm\mu_l}{\Gamma(1+1/\kappa)}$, $S_l(t) = \exp\{-(t/\bm\lambda_l)^\kappa\}$, $\forall l$
      \State Randomly select one $l$ from $\{1,...,L\}$, and 
      draw $(\bm\beta_l,\bm\gamma_l)$ via M-H sampling
      \State Draw all $\bm\beta_l$ via ARMS, $\forall l$ 
      \State Update Weibull's mean parameter $\mu_l=\exp\{\beta_{0l}+\mathbf X_l\bm\beta_l\}$, and update $\bm\lambda_l$ and $S_l(t)$
  \EndFor
  \end{algorithmic}
  }
\end{algorithm}
\medskip
{\noindent\bf Specification of hyperparameters}
\smallskip\\
For the Weibull's shape parameter, we use a non-informative prior $\kappa\sim\mathcal Gamma(1, 1)$. 
As a practical suggestion, all the covariates can be standardized, i.e., transforming to $z$-score values. 
Since a weakly informative inverse-gamma prior is used for each variance (i.e., $v_0^2, v^2, \tau_0^2, \tau_l^2, w_0^2, w_l^2$), we set a prior $\mathcal{IG}amma(5, 20)$ by default. 
If the magnitude of some covariates is very large, the inverse-gamma prior can be specified with a large variance. 
For GPTCM-Ber1 and GPTCM-Ber2 with the Bernoulli-beta prior, the Bernoulli's probability is assigned by a beta prior $\mathcal Beta(1, cp)$, where $p$ is the number of cell-type-specific covariates, and $c>0$ is a tuning parameter. 
For GPTCM-MRF1 and GPTCM-MRF2 with a MRF prior, the sparsity parameter (i.e., $a$ or $a^\star$) is chosen according to the logistic transformation with an assumed model sparsity $s$, i.e., $a=\text{logit}(s)$, since the MRF prior is reduced to independent Bernoulli priors with probability logit$^{-1}(s)$ if $b=0$. 
The MRF hyperparameter $b$ or $b^\star$ for the strength of the structure relationships between variables is tuned via a grid search. 
The graph matrix ($G$ or $G^\star$) in the MRF prior is a weight matrix with weight $1$ between two linked variables in a given biological network/pathway and weight $0.5$ (or any value between $0$ and $1$) between two variables that are naturally correlated (e.g. the same gene with different representations in different cell types). 
For the tuning parameters, their optimal values can be chosen based on the criteria of the expected log pointwise predictive density (elpd) calculated by the approximate leave-one-out cross-validation \citep{Vehtari2024}. 

\section{Simulations}\label{sec:simulations}

We use simulations to gain insight into the performance of the proposed Bayesian GPTCMs, as well as comparisons with classical survival methods.  
We simulate $L=3$ cell types with both low-dimensional ($p=10, n=200$) and high-dimensional cell-type-specific covariates ($p=200, n=200$) $\mathbf X_l\in \mathbb R^{n\times p}$, $l=1,...,L$, and assume only a few covariates are truly relevant. 
The cure rate parameter is modeled by two mandatory covariates $\mathbf X_0=\{x_{(0)i1},x_{(0)i1},\}_{i=1}^n$ via a logarithmic link function. 
Censoring is generated through an exponential distribution that results in approximately 20\% censoring rate. 
See below the simulation scheme: 
\begin{align*}
  x_{(0)i1} &\sim \mathcal Bern(0.5),\ x_{(0)i2} \sim \mathcal N(0,1), \ \ i=1,...,n \\
  \theta_i &= \exp\left\{1+0.6x_{(0)i1}-x_{(0)i2}\right\}\\
  \text{vec}[(\mathbf X_{i1},...,\mathbf X_{iL})] &\sim \mathcal N(0, \Sigma),\ \ \mathbf X_{il}=(x_{i1l},...,x_{ipl})^\top,\ l=1,...,L\\
  \mu_{il} &= \exp\{\mathbf X_{il} \bm\beta_l\} \\
  \alpha_{il} &= \exp\{\zeta_{0l} + \mathbf X_{il} \bm\zeta_l\} \\ 
  \mathbf p_{i\cdot} = (\mathtt p_{i1},...,\mathtt p_{iL}) &\sim \mathcal Dirichlet(\alpha_{i1},...,\alpha_{iL}) \\
  C_i^1 &\sim \mathcal Uniform(1,4) \\
  C_i^2 &\sim \mathcal Exponential(-\log(0.8)/5) \\
  C_i &= \min\{C_i^1, C_i^2\} \\
  U_{i} &\sim \mathcal Uniform(0, 1) \\
  T_i &= 
  \begin{cases}
    +\infty, & U_{i} \le e^{-\theta_i}\\
    \text{Metropolis-Hastings sampler for }S_{pop}(t), & U_{i} > e^{-\theta_i}
  \end{cases}\\
  t_i &= \min\{T_i, C_i\}\\
  \delta_i &= \mathbbm 1\{T_i \le C_i\}.
\end{align*}
Note that to simulate $U_i$ above is for generating the cure population. 
If $U_{i} \le e^{-\theta_i}$, then the $i$-th subject was censored, 
so the observed time $t_i$ is equal to the censored time $C_i$ smaller than the true survival time, i.e., $t_i=\min\{T_i, C_i\}=C_i$, and no event happened $\delta_i=0$. 
The Weibull's shape parameter is set as $\kappa=2$. 
The covariance matrix of the cell-type-specific covariates is set as
\[
\Sigma = 
\begin{pmatrix}
  \Sigma_1    & \varrho\mathbb I_p & \cdots & \varrho\mathbb I_p \\
  \varrho\mathbb I_p & \Sigma_2    & \cdots & \varrho\mathbb I_p \\
  \vdots      & \vdots      & \ddots & \vdots \\
  \varrho\mathbb I_p & \varrho\mathbb I_p & \cdots & \Sigma_L
\end{pmatrix}.
\]
The diagonals of the covariance matrix $\Sigma$ are 1. 
In the off-diagonal blocks of $\Sigma$, $\varrho=0.1$ meaning that the correlation between the same gene with different gene expression levels from any two cell types is $0.1$. 
In the $l$-th block diagonal matrix, $\Sigma_l=\{\sigma_{jj'l}\}_{jj'}$ with $\sigma_{jj'l}=\varrho_l^{|j-j'|}$ if $j, j'\le 6$ and $\sigma_{jj'l}=0$ otherwise, $\varrho_1=0.13$, $\varrho_2=0.14$ and $\varrho_3=0.15$. 
The effects of cell-type-specific covariates are set as 
\begin{align*}
  \begin{array}{*{24}{c@{\hspace{4pt}}}}
\bm\beta_1 &= &(&-1.0, &-0.5, &0.8, &0.8, &-1.0, &0, &0,\ \ \ \ \underbrace{0, \cdots, 0}_{\# p-7}\ \ )^\top, \\
\bm\beta_2 &= &(&0, &-0.9, &-0.8, &0, &1.5, &1, &0,\ \ \ \ \underbrace{0, \cdots, 0}_{\# p-7}\ \ )^\top, \\
\bm\beta_3 &= &(&1.0, &0, &-0.4, &-1.5, &0, &0, &0.8,\ \ \ \ \underbrace{0, \cdots, 0}_{\# p-7}\ \ )^\top, \\
\bm\zeta_1 &= &(&0.7, &-0.7, &0.5, &-0.5, &1, &0, &0,\ \ \ \ \underbrace{0, \cdots, 0}_{\# p-7}\ \ )^\top, \\
\bm\zeta_2 &= &(&-0.5, &0.5, &0, &1, &0, &-1, &0,\ \ \ \ \underbrace{0, \cdots, 0}_{\# p-7}\ \ )^\top, \\
\bm\zeta_3 &= &(&0, &0, &1, &-0.5, &-0.7, &0, &0,\ \ \ \ \underbrace{0, \cdots, 0}_{\# p-7}\ \ )^\top.
  \end{array}
\end{align*}
The intercepts in the measurement error submodel for the Dirichlet concentration parameters are set as 
$(\bm\zeta_{01},\bm\zeta_{02},\bm\zeta_{03}) = (-.5,1,1.2)$. 
We simulate additionally independent $n=200$ samples as validation data to assess the prediction performance of all models. 

Furthermore, we simulate data from a Cox proportional hazards model to assess the performance of the proposed Bayesian GPTCMs under model misspecification. 
Five covariates are simulated independently from a standard normal distribution, with the number of subjects $n=200$. 
The effects of the five covariates are set as $(-0.8, -2, -2, 1, 1)$. 
The survival data are simulated based on the Cox-Weibull model \citep{Bender2005} with the Weibull's shape parameter $\kappa=2$ and the censoring rate is controlled as approximately 20\%. 

\subsection{Simulation results in low dimensions} \label{sec:simLow}

We compare the six Bayesian versions of the GPTCM proposed in Section \ref{sec:methods} as well as the classical Cox model \citep{Cox1972} and the semiparametric promotion time cure model (PTCM) \citep{Ma2008}. 
Note that the Cox model and PTCM are not suited to model the cell-type proportions data. 
We implemented three versions of the Cox model: (i) ``Cox-X0'': with only clinical covariates $\mathbf X_0$, (ii) ``Cox-Xmean'': with mean aggregate covariates of $\mathbf X_1,...,\mathbf X_L$, i.e., each covariate is the mean of the corresponding $L$ cell-type-specific covariates, and (iii) ``Cox-X0-Xmean'': with both covariates from (i) and (ii). 
For the semiparametric PTCM, only the clinical variables were included for modeling the cure fraction, denoted as ``PTCM-X0''. 
For the GPTCMs with a MRF prior (i.e., GTPCM-MRF1 and GPTCM-MRF2), a graph based on the precision matrix $\Sigma^{-1}$ was used, which has weights 1 corresponding to nonzero entries of the precision matrix but weights 0.5 corresponding to the same variable with $L$ representative features in different cell types. 
For each of the six GPTCMs, we ran 25,000 MCMC iterations with the first 5,000 iterations as a warmup period. 

\begin{figure}[!htbp]
  \centering
  \includegraphics[height=0.35 \textwidth]{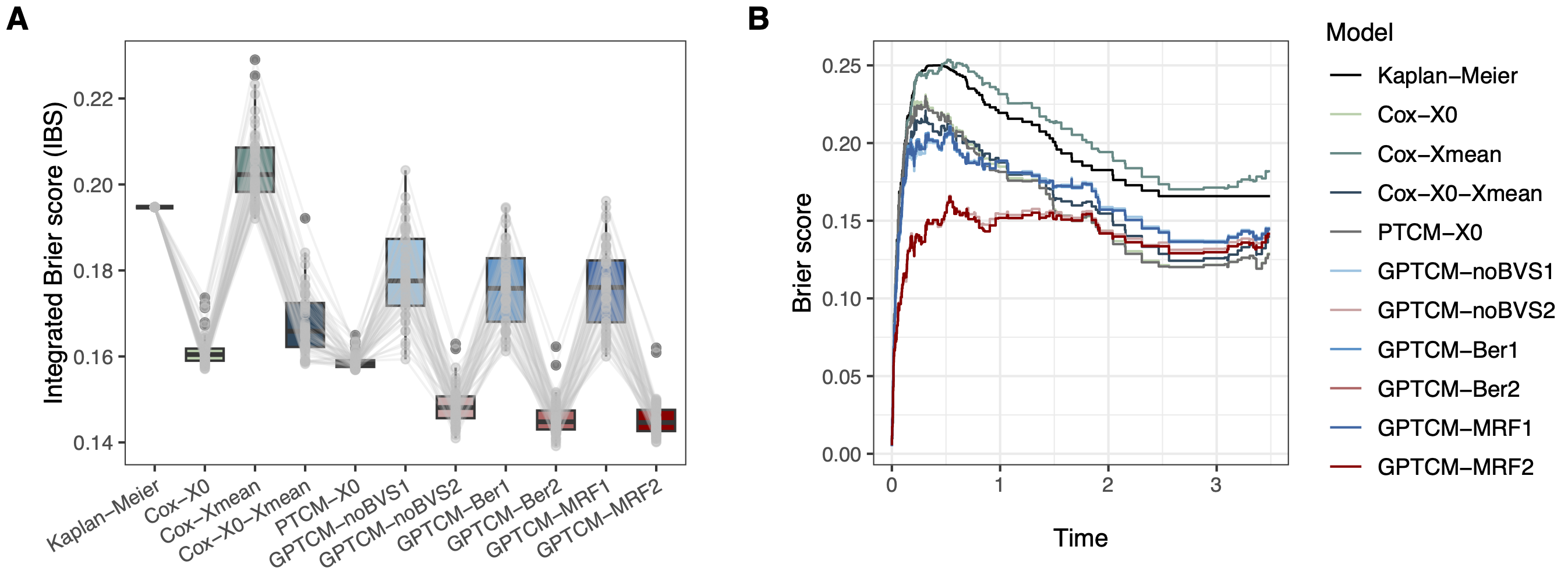}
  \caption{\it Simulation results in low dimensions: Out-sample prediction errors of classical survival models and GPTCMs in terms of integrated Brier score (IBS) trained by 50 simulated data sets (panel A) and time-dependent Brier score trained by one simulated data set (panel B). 
  The grey colored lines in panel A connect the performances of different models on the same data set. 
  The Kaplan-Meier method is as a reference that did not account for any covariate. 
  The method ``Cox-X0'' only models clinical covariates $\mathbf X_0$. 
  The method ``Cox-Xmean'' models mean aggregate covariates of $\mathbf X_1,...,\mathbf X_L$. 
  The method ``Cox-X0-Xmean'' models both clinical covariates $\mathbf X_0$ and mean aggregate covariates of $\mathbf X_1,...,\mathbf X_L$. 
  The predictions of GPTCM-noBVS1 and GPTCM-noBVS2 were based on the posterior mean of all parameters. The predictions of other GPTCMs were based on the estimates of their median probability models (MPMs) for the effects of covariates with variable selection and estimates of posterior mean for other parameters.}
  \label{fig:simLow-brier}
\end{figure}

Figure \ref{fig:simLow-brier} shows the survival prediction in terms of integrated Brier score (IBS) and time-dependent Brier score by all models based on a validation data set, 
which indicates the superior prediction performances of three GPTCMs (i.e., GPTCM-noBVS2, GPTCM-Ber2 and GPTCM-MRF2) that model covariates for both cell-type-specific survival and cell-type-specific proportions. 
The time-dependent Brier score in Figure \ref{fig:simLow-brier}B shows that the performances of three GPTCMs without modeling covariates for cell-type-specific proportions (i.e., GPTCM-noBVS1, GPTCM-Ber1 and GPTCM-MRF1) perform similarly to the classical models Cox-X0, Cox-X0-Xmean and PTCM-X0 for short-term survival predictions, but they become worse than those classical models for long-term survival predictions. 
The Cox model with only mean aggregate covariates omitting clinical covariates (i.e., Cox-Xmean) has worse prediction performance than the Kaplan-Meier method.  

\begin{figure}[!htbp]
  \centering
  \makebox[\textwidth][c]{
  \includegraphics[height=0.8 \textwidth]{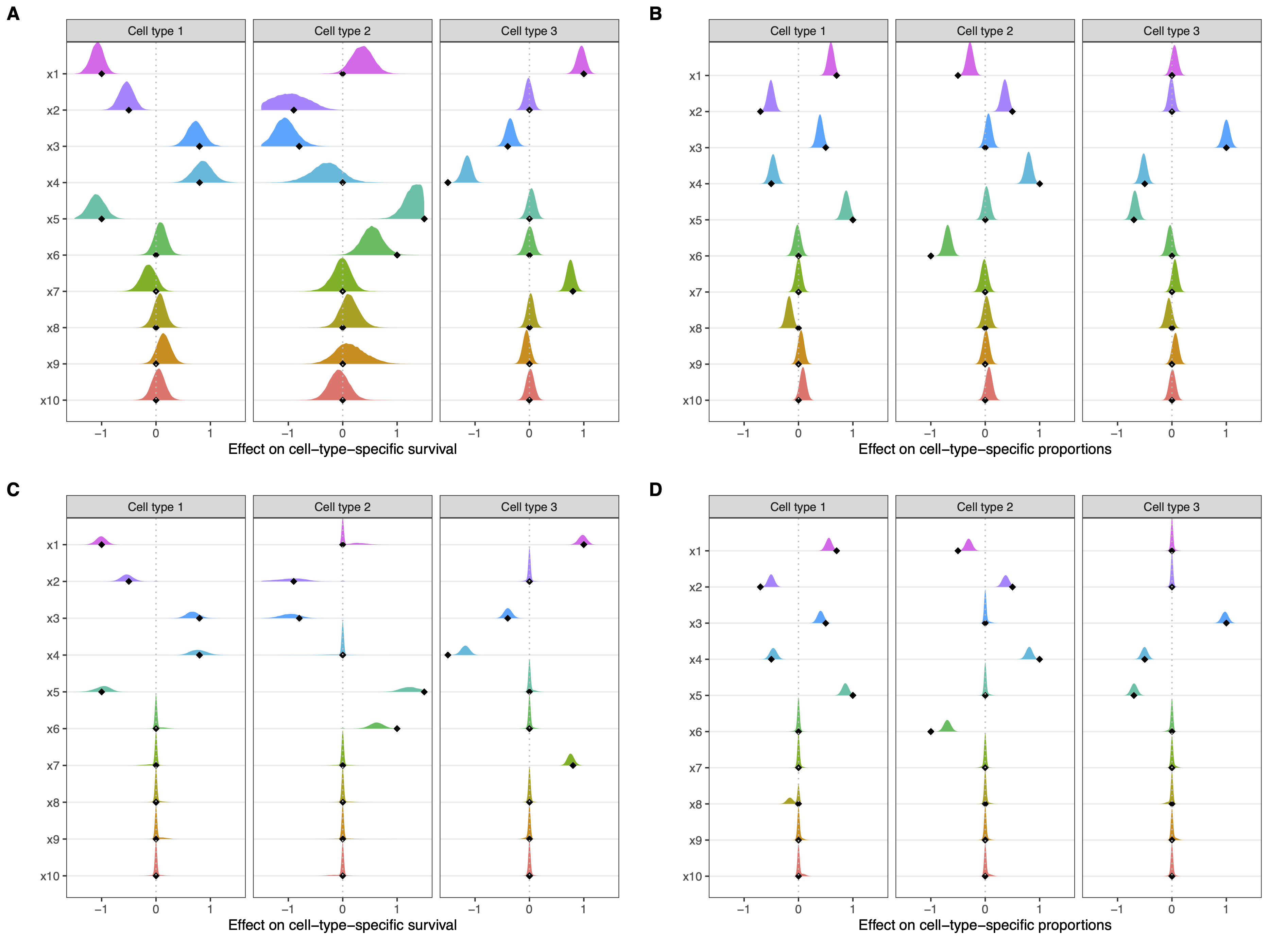} }
  \caption{\it Simulation results in low dimensions: Posterior distributions of the effects on cell-type-specific survival and proportions based on one simulated data set. The black colored diamond indicates the true effect. (A) Posterior distributions of the effects on cell-type-specific survival by GPTCM-noBVS2. (B) Posterior distributions of the effects on cell-type-specific proportions by GPTCM-noBVS2. (C) Posterior distributions of the effects on cell-type-specific survival by GPTCM-MRF2. (D) Posterior distributions of the effects on cell-type-specific proportions by GPTCM-MRF2.}
  \label{fig:simLow-postCoeff}
\end{figure}

Figure \ref{fig:simLow-postCoeff} shows the posterior distributions of the effects on cell-type-specific survival and effects on cell-type-specific proportions by GPTCM-noBVS2 (Figure \ref{fig:simLow-postCoeff}A-B) and GPTCM-MRF2 (Figure \ref{fig:simLow-postCoeff}C-D). 
Although all the posterior distributions of the effects from GPTCM-noBVS2 
(Figure \ref{fig:simLow-postCoeff}A-B) 
cover the true effects almost, they have much larger variations than the posterior distributions from GPTCM-MRF2 
(Figure \ref{fig:simLow-postCoeff}C-D). 
In particular, GPTCM-MRF2 results in the estimates of the truly unrelated variables (true effects as zero) with very high densities at the value zero and the estimates of the truly related variables with most densities far away from the value zero, i.e., with accurate variable selection. 
Supplementary Figure \ref{figS:simLow-mPIP}A-B also shows that GPTCM-MRF2 results in the marginal posterior inclusion probabilities (mPIPs) of the truly unrelated variables close to 0 and the mPIPs of the truly related variables close to 1. 
GPTCM-Ber2 has similar performance to GPTCM-MRF2 in terms of variable selection and effect estimation, see Supplementary Figure \ref{figS:simLow-mPIP}C-D and Supplementary Figure \ref{figS:simLow-postCoeff-gptcm4}. 
For other three GPTCMs with only modeling covariates for cell-type-specific survival (i.e., GPTCM-noBVS1, GPTCM-Ber1 and GPTCM-MRF1), GPTCM-noBVS1 performs similarly to GPTCM-noBVS2 in terms of the effects on cell-type-specific survival (Supplementary Figure \ref{figS:simLow-postCoeff-gptcm135}A), and GPTCM-Ber1 and GPTCM-MRF1 
perform similarly to GPTCM-Ber2 and GPTCM-MRF2 in terms of the posterior distributions of the effects on cell-type-specific survival (Supplementary Figure \ref{figS:simLow-postCoeff-gptcm135}B-C). 

\begin{table}[!htbp]
\centering
\caption{Simulation results in low dimensions: Model estimation and variable selection\medskip\label{tab:simLow} }
\medskip
\resizebox{\textwidth}{!}{
\begin{tabular}{lcccccccccc}  
\toprule
  \multirow{2}{*}{\bf Model} & 
  \multirow{2}{*}{$\frac{1}{\sqrt{pL}}\|\hat{\bm\beta}-\bm\beta\|_2$} & 
  \multirow{2}{*}{$\frac{1}{\sqrt{pL}}\|\hat{\bm\zeta}-\bm\zeta\|_2$} & 
  \multicolumn{2}{c}{\bf Accuracy}  & & 
  \multicolumn{2}{c}{\bf Sensitivity} & &  
  \multicolumn{2}{c}{\bf Specificity}  \\
\cmidrule{4-5}
\cmidrule{7-8}
\cmidrule{10-11}
 & & & $\bm\gamma$ & $\bm\eta$ && $\bm\gamma$ & $\bm\eta$ && $\bm\gamma$ & $\bm\eta$ \\
\midrule
GPTCM-noBVS1      & 0.196 (0.0920) &       & \\
GPTCM-noBVS2      & 0.226 (0.0765) & 0.111 (0.0131) & \\
GPTCM-Ber1        & 0.141 (0.0541) &                & 0.979 (0.0276) &                && 0.968 (0.0442) &       && 0.987 (0.0321) &        \\
GPTCM-Ber2        & 0.168 (0.0686) & 0.101 (0.0134) & 0.960 (0.0369) & 0.973 (0.0294) && 0.938 (0.0581) & 1.000 (0.0000) && 0.976 (0.0376) & 0.956 (0.0489)  \\
GPTCM-MRF1        & 0.139 (0.0541) &                & 0.973 (0.0343) &                && 0.977 (0.0418) &       && 0.971 (0.0449) &        \\
GPTCM-MRF2        & 0.169 (0.0716) & 0.099 (0.0140) & 0.962 (0.0343) & 0.985 (0.0262) && 0.932 (0.0615) & 1.000 (0.0000) && 0.985 (0.0310) & 0.976 (0.0437)  \\
\bottomrule
\multicolumn{11}{p{240mm}}{NOTE:\it Each value outside parentheses is the mean estimate over 50 repeated simulations, and a value inside parentheses is the standard deviation of the estimate over 50 repeated simulations. 
 The empty cells are due to no parameters in corresponding models. 
 The variable selection performance (i.e., accuracy, sensitivity and specificity) of the last four GPTCMs (i.e., GPTCM-Ber1, GPTCM-Ber2, GPTCM-MRF1 and GPTCM-MRF2) is determined by the median probability model with mPIP thresholded at 0.5. 
 } 
\end{tabular}
}
\end{table}

Table \ref{tab:simLow} shows that the four GPTCMs with BVS (i.e., GPTCM-Ber1, GPTCM-Ber2, GPTCM-MRF1 and GPTCM-MRF2) outperform GPTCM-noBVS1 and GPTCM-noBVS2 
in terms of estimated effects on the cell-type-specific survival, i.e., $\frac{1}{\sqrt{pL}}\|\hat{\bm\beta}-\bm\beta\|_2$, also see Supplementary Figure \ref{figS:simLow-rmseBeta}A. 
In terms of the estimated effects on the cell-type-specific proportions, $\frac{1}{\sqrt{pL}}\|\hat{\bm\zeta}-\bm\zeta\|_2$, the two GPTCMs with BVS (i.e., GPTCM-Ber2 and GPTCM-MRF2) also outperform GPTCM-noBVS2, also see Supplementary Figure \ref{figS:simLow-rmseBeta}B.  
All the four GPTCMs with BVS (i.e., GPTCM-Ber1, GPTCM-Ber2, GPTCM-MRF1 and GPTCM-MRF2) based on their MPMs with mPIPs thresholded at $0.5$ have well identified truly related and truly unrelated variables, see the variable selection accuracy, sensitivity and specificity in Table \ref{tab:simLow}. 

\begin{figure}[!htbp]
  \centering
  \includegraphics[height=0.35 \textwidth]{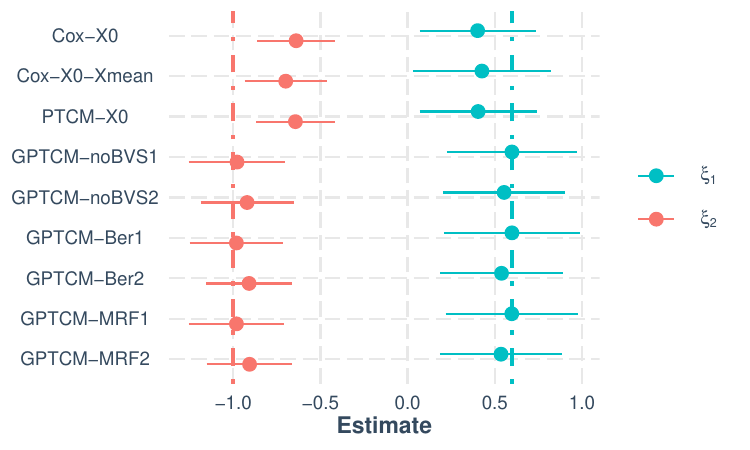} 
  \caption{\it Simulation results in high dimensions: Estimates of the clinical variables' effects on the cure fraction. The solid circle point shows a mean estimate over 50 repeated simulations. The error bar shows
  a $95\%$ confidence interval produced by the 50 repeated simulations. The green and red colored dot-dashed lines denote the true effects of the two clinical variables, respectively. 
  }
  \label{fig:simLow-xi}
\end{figure}

For the effect estimation of clinical variables, Figure \ref{fig:simLow-xi} shows that all the six GPTCMs resulted in good estimates of the two clinical variables' effects. 
However, the three classical survival models (i.e., Cox-X0, Cox-X0-Xmean and PTCM-X0) resulted in more biased estimates for the first clinical variable's effect $\xi_1$, particularly bad estimates for the second clinical variable's effect $\xi_2$. 

In addition, we simulated data based on a Cox proportional hazards model to assess the performance of GPTCMs under model misspecification. 
Since the underlying true model only includes five clinical variables, we fitted GPTCMs by including the five variables in both cure and noncure fractions. 
Since GPTCM-Ber2 and GPTCM-MRF2 also need proportions data as input data, we used pseudo proportions data of three cell types with all values $1/3$. 
The MRF prior uses a graph matrix with nonzero weights 1 only for the same variable across the three pseudo cell types.  
Supplementary Figure \ref{figS:simLow-brier-misspecification} shows that the two classical survival models (i.e., Cox-X0, PTCM-X0) have the best prediction performances in terms of time-dependent Brier score. 
The four GPTCMs with BVS (i.e., GPTCM-Ber1, GPTCM-Ber2, GPTCM-MRF1 and GPTCM-MRF2) perform slightly worse than the classical survival models for short-term survival prediction, but their performances become much worse for long-term survival prediction. 
The other two GPTCMs without variable selection (i.e., GPTCM-noBVS1 and GPTCM-noBVS2) have much worse survival prediction than the Kaplan-Meier method (Supplementary Figure \ref{figS:simLow-brier-misspecification}). 
From Supplementary Figure \ref{figS:simLow-postCoeff-misspecification}A-C, we see that the cell-type-specific effect estimates by GPTCM-noBVS1 and GPTCM-noBVS2 have very large variation. 
In contrast, the other four GPTCMs with BVS (i.e., GPTCM-Ber1, GPTCM-Ber2, GPTCM-MRF1 and GPTCM-MRF2) estimated all cell-type-specific effects with accurate value $0$ as expected (Supplementary Figure \ref{figS:simLow-postCoeff-misspecification}D-I). 
For the effect estimation of the clinical covariates in the cure fraction, GPTCM-noBVS1 and GPTCM-noBVS2 result in largely variational estimated posterior distributions, with all $95\%$ credible intervals covering the value 0. 
However, GPTCM-MRF1 and GPTCM-MRF2 result in estimates with narrow $95\%$ credible intervals that cover the values of true effects and far away from $0$ similar to the Cox model under the correct model specification (Supplementary Figure \ref{figS:simLow-xi-misspecification}). 

\subsection{Simulation results in high dimensions}

Since GPTCM-noBVS1, GPTCM-noBVS2 and classical survival models cannot deal with high-dimensional covariates, we compare the other four GPTCMs (i.e., GPTCM-Ber1, GPTCM-Ber2, GPTCM-MRF1 and GPTCM-MRF2), the classical Cox model with only clinical covariates, and two elastic net Cox models \citep{Simon2011}. 
Both the two elastic net Cox models include the clinical covariates as mandatory variables and perform variable selection for other high-dimensional covariates with an $\ell_1/\ell_2$-norm (or elastic net) penalty. 
The first elastic net Cox model (denoted as ``Enet Cox1'') performs variable selection for the $p$ mean aggregate covariates of $\mathbf X_1,...,\mathbf X_L$, but the second elastic net Cox model (denoted as ``Enet Cox2'') performs variable selection for all $pL$ individual cell-type-specific covariates $[\mathbf X_1,...,\mathbf X_L]$. 
The elastic net penalty parameters were optimized by using $5$-fold cross-validation based on a training data set. 
For the GPTCMs with a MRF prior (i.e., GTPCM-MRF1 and GPTCM-MRF2), a graph based on the precision matrix $\Sigma^{-1}$ was used in the same way as the setting in low dimensions in Section \ref{sec:simLow}. 
For each of the four GPTCMs, we ran 500,000 MCMC iterations with the first 200,000 iterations as a warmup period. 

\begin{figure}[!htbp]
  \centering
  \includegraphics[height=0.36 \textwidth]{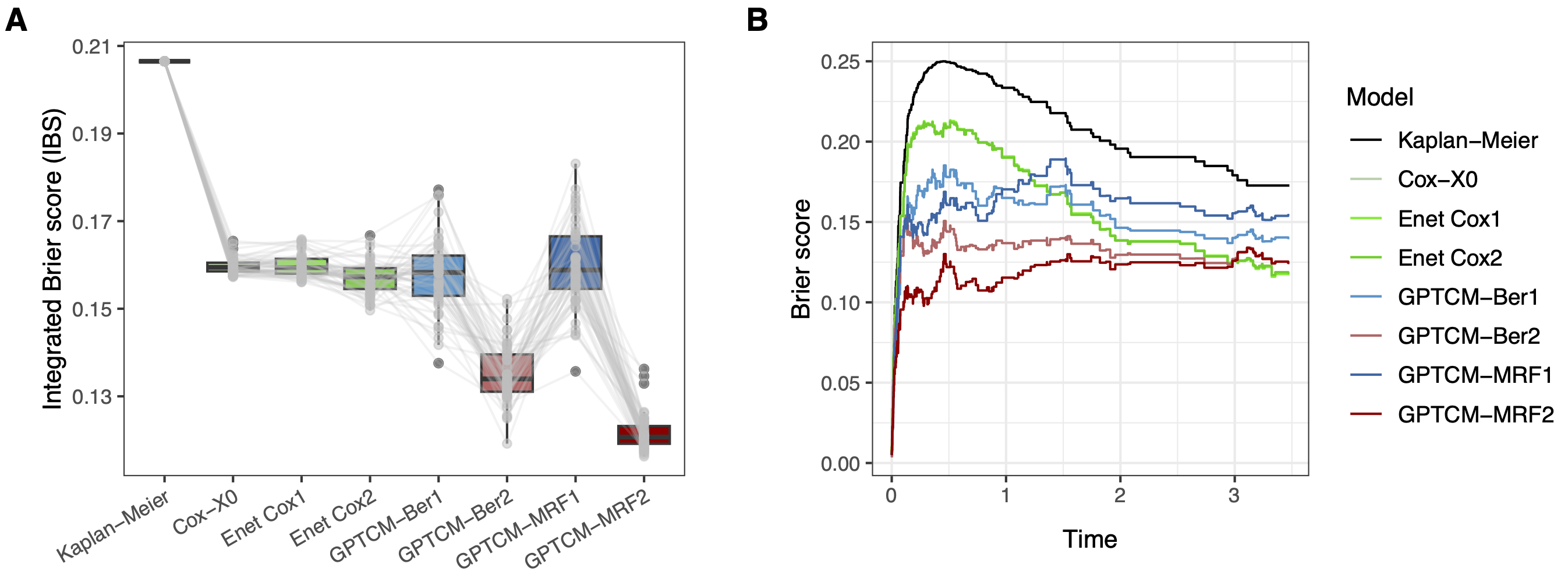}
  \caption{\it Simulation results in high dimensions: 
  Out-sample prediction errors of classical survival models and GPTCMs in terms of integrated Brier score (IBS) trained by 50 simulated data sets (panel A) and time-dependent Brier score trained by one simulated data set (panel B). 
  The grey colored lines in panel A connect the performances of different models on the same data set. 
  The Kaplan-Meier method is as a reference that did not account for any covariate. 
  The elastic net Cox model ``Enet Cox1'' included the clinical variables $\mathbf X_0$ and mean aggregate variables of $\mathbf X_1,...,\mathbf X_L$.  The elastic net Cox model ``Enet Cox2'' included the clinical variables $\mathbf X_0$ and all cell-type-specific variables $[\mathbf X_1,...,\mathbf X_L]$. 
  The predictions of other GPTCMs were based on the estimates of their median probability models (MPMs) for the effects of covariates with variable selection and estimates of posterior mean for other parameters. }
  \label{fig:simHigh-brier}
\end{figure}

Figure \ref{fig:simHigh-brier} shows the out-sample survival prediction in terms of integrated Brier score (IBS) and time-dependent Brier score by all models based on one validation data set, 
which indicates the superior prediction performances of two GPTCMs (i.e., GPTCM-Ber2 and GPTCM-MRF2) that model covariates for both cell-type-specific survival and cell-type-specific proportions, in which GPTCM-MRF2 with a prior graph outperforms GPTCM-Ber2. 
The performances of the other two GPTCMs (i.e., GPTCM-Ber1 and GPTCM-MRF1) without modeling covariates for cell-type-specific proportions outperform the two elastic net models for short-term survival predictions (until time point $\sim$1.2, Figure \ref{figS:simHigh-brier1.2}), but they lead to worse long-term survival predictions than the classical Cox model, Enet Cox1 and Enet Cox2 (Figure \ref{fig:simHigh-brier}B).  

\begin{figure}[!htbp]
  \centering
  \makebox[\textwidth][c]{
  \includegraphics[height=0.9 \textwidth]{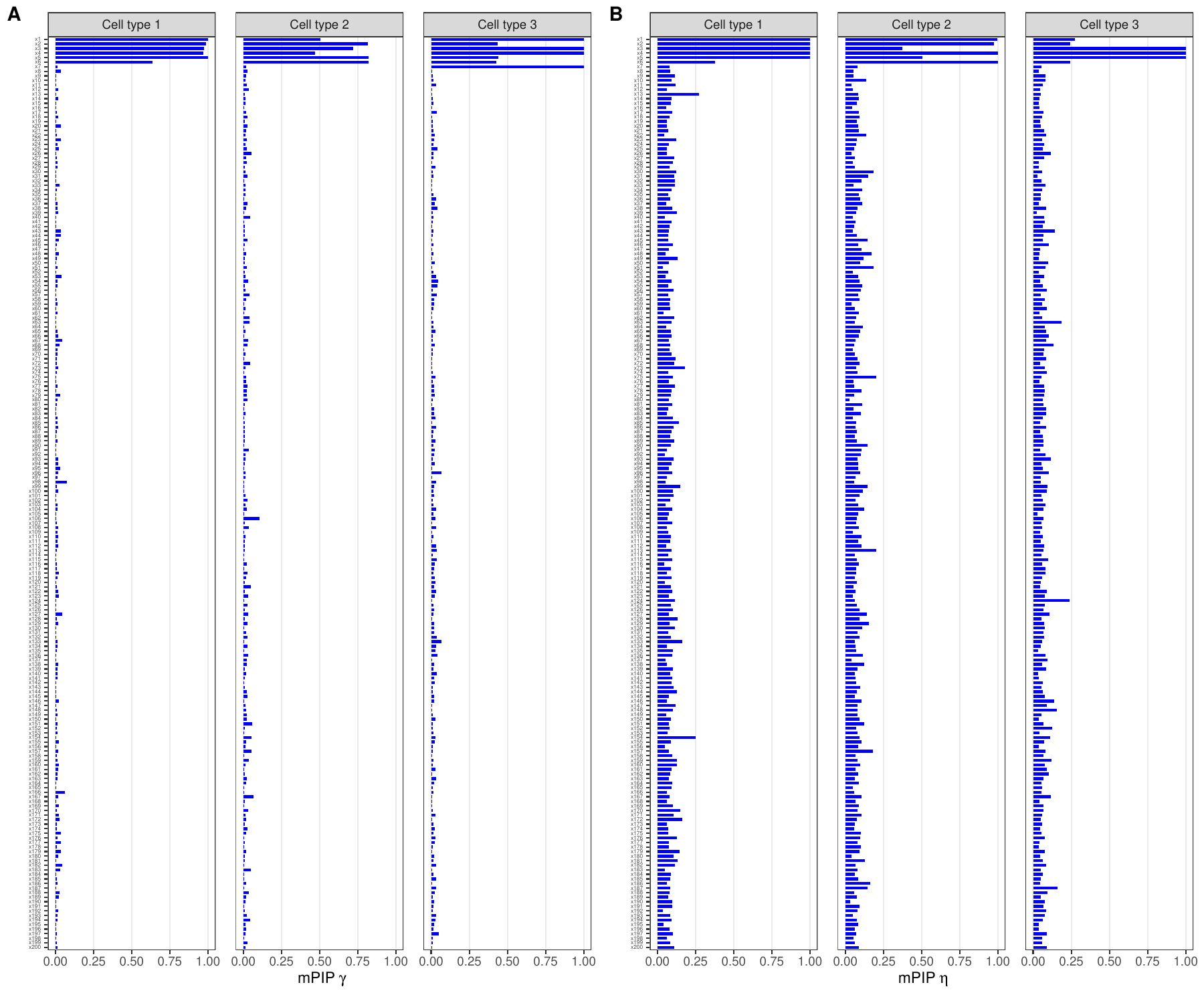} }
  \caption{\it Simulation results in high dimensions: Marginal posterior inclusion probabilities (mPIPs) of the variables linked to cell-type-specific survival and proportions by GPTCM-MRF2 based on one simulated data set. (A) mPIPs of the variables linked to cell-type-specific survival. (B) mPIPs of the variables linked to cell-type-specific proportions. 
  }
  \label{fig:simHigh-mPIP-gptcm6}
\end{figure}

Although the Enet Cox2 can induce variable selection from all individual cell-type-specific covariates, it only identified $21.5\%$ (standard deviation $19.47\%$) of truly related covariates (Table \ref{tab:simHigh}). 
Table \ref{tab:simHigh} shows that GPTCM-Ber2 without prior knowledge results in good variable selection linked to the cell-type-specific proportions (see sensitivity [$0.992 (0.0304)$] and specificity [$1.000 (0.0003)$] of $\bm\eta$ in Table \ref{tab:simHigh}), but GPTCM-Ber2 did not identify many truly related covariates linked to the cell-type-specific survival (see sensitivity [$0.349 (0.1090)$] of $\bm\gamma$ in Table \ref{tab:simHigh}). 
Supplementary Figure \ref{figS:simHigh-postCoeff-gptcm135}A-B show that GPTCM-Ber1 and GPTCM-Ber2 cannot capture a few nonzero effects on cell-type-specific survival corresponding to cell type 1 and cell type2, but GPTCM-MRF1 and GPTCM-MRF2 have better estimation (Supplementary Figure \ref{figS:simHigh-postCoeff-gptcm135}D-E). 
GPTCM-MRF2 with prior graph knowledge of cell-type-specific covariates can correctly identify most truly related covariates (see variable selection sensitivities of $\bm\gamma$ [$0.931 (0.1460)$] and $\bm\eta$ [$1.000 (0.0000)$] in Table \ref{tab:simHigh}). 
Figure \ref{fig:simHigh-mPIP-gptcm6} shows that variable selection of GPTCM-MRF2 based on one simulated data set, with large mPIPs for truly relevant covariates among the first a few variables and with very small mPIPs for other irrelevant covariates.

\begin{table}[!htbp]
\centering
\caption{Simulation results in high dimensions: Model estimation and variable selection\medskip\label{tab:simHigh} }
\medskip
\resizebox{\textwidth}{!}{
\begin{tabular}{lcccccccccc}  
\toprule
  \multirow{2}{*}{\bf Model} & 
  \multirow{2}{*}{$\frac{1}{\sqrt{pL}}\|\hat{\bm\beta}-\bm\beta\|_2$} & 
  \multirow{2}{*}{$\frac{1}{\sqrt{pL}}\|\hat{\bm\zeta}-\bm\zeta\|_2$} & 
  \multicolumn{2}{c}{\bf Accuracy}  & & 
  \multicolumn{2}{c}{\bf Sensitivity} & &  
  \multicolumn{2}{c}{\bf Specificity}  \\
\cmidrule{4-5}
\cmidrule{7-8}
\cmidrule{10-11}
 & & & $\bm\gamma$ & $\bm\eta$ && $\bm\gamma$ & $\bm\eta$ && $\bm\gamma$ & $\bm\eta$ \\
\midrule
Enet Cox1      & 0.144 (0.0016) &                & 0.793 (0.3422) &                && 0.257 (0.3491) &       && 0.805 (0.3572) &        \\
Enet Cox2      & 0.146 (0.0027) &                & 0.927 (0.1873) &                && 0.215 (0.1947) &       && 0.942 (0.1951) &        \\
GPTCM-Ber1        & 0.105 (0.0197) &                & 0.987 (0.0037) &                && 0.462 (0.1569) &       && 0.999 (0.0012) &        \\
GPTCM-Ber2        & 0.114 (0.0158) & 0.023 (0.0046) & 0.985 (0.0025) & 1.000 (0.0007) && 0.349 (0.1090) & 0.992 (0.0304) && 1.000 (0.0008) & 1.000 (0.0003)  \\
GPTCM-MRF1        & 0.039 (0.0214) &                & 0.993 (0.0023) &                && 0.962 (0.0961) &       && 0.994 (0.0027) &        \\
GPTCM-MRF2        & 0.045 (0.0263) & 0.020 (0.0028) & 0.993 (0.0023) & 0.999 (0.0011) && 0.931 (0.1460) & 1.000 (0.0000) && 0.994 (0.0026) & 0.999 (0.0011)  \\
\bottomrule
\multicolumn{11}{p{250mm}}{NOTE:\it Each value outside parentheses is the mean estimate over 50 repeated simulations, and a value inside parentheses is the standard deviation of the estimate over 50 repeated simulations. 
The empty cells are due to no parameters in corresponding models. 
The variable selection of the GPTCMs is determined by the median probability model (MPM) with marginal posterior inclusion probability (mPIP) thresholded at 0.5. 
} 
\end{tabular}
}
\end{table}

\begin{figure}[!htbp]
  \centering
  \includegraphics[height=0.35 \textwidth]{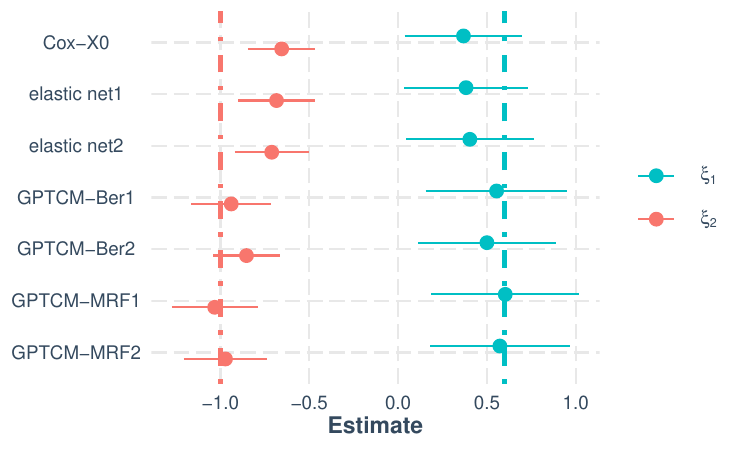} 
  \caption{\it Simulation results in high dimensions: Estimates of the clinical variables' effects on the cure fraction. The solid circle point shows a mean estimate over 50 repeated simulations. The error bar shows
  a $95\%$ confidence interval produced by the 50 repeated simulations. The green and red colored dot-dashed lines denote the true effects of the two clinical variables, respectively. 
  }
  \label{fig:simHigh-xi}
\end{figure}

For the effect estimation of clinical variables, Figure \ref{fig:simHigh-xi} shows that all the four GPTCMs resulted in well estimated effects with their $95\%$ confidence intervals covering the true effect values. 
However, the classical Cox model, Enet Cox1 and Enet Cox2 result in biased effect estimation of $\xi_2$ for the second clinical variable.

\section{Discussion and conclusion}\label{sec:discussion}

We have presented a class of six versions of a Bayesian generalized promotion time cure model (GPTCM) for both low-dimensional and high-dimensional data. 
The proposed Bayesian hierarchical modeling framework intuitively integrates multiscale data including individual-level survival data, multicellular-level cell type proportions and cell-type-specific covariates.  
Four of the GTPCMs (i.e., GPTCM-Ber1, GPTCM-Ber2, GPTCM-MRF1 and GPTCM-MRF2) allow variable selection for cell-type-specific covariates, in which GPTCM-Ber2 and GPTCM-MRF2 allow both, variable selection for covariates linked to cell-type-specific survival and variable selection for covariates linked to cell-type-specific proportions. 
A unified full Bayesian inference has been provided for the computations of all the GPTCMs. 
In particular, the GPTCM with a MRF prior can make use of known biological network information between (cell-type-specific) covariates. 
When biological research advances, the GPTCM with a MRF prior could improve just from the improved network information provided, without any additional training data or improved methodology. 

The simulation studies indicate that the GPTCMs with variable selection are useful when unrelated covariates are present in both low- and high-dimensional data for survival prognosis (Figures \ref{fig:simLow-brier} and \ref{fig:simHigh-brier}). 
Note that even in simulation scenarios where the data were generated such that the GPTCM models were misspecified, i.e. there was no cure fraction and the true data-generating process corresponded to a Cox proportional hazards model with only a few mandatory clinical variables, the GPTCMs with variable selection (i.e., GPTCM-Ber1, GPTCM-Ber2, GPTCM-MRF1 and GPTCM-MRF2) showed good performance. These models performed only slightly worse in terms of survival prognosis than the (correctly specified) Cox model (Supplementary Figure \ref{figS:simLow-postCoeff-misspecification}), but the four GPTCMS could correctly identify unrelated covariates (Supplementary Figure \ref{figS:simLow-postCoeff-misspecification}), and GPTCM-MRF1 and GPTCM-MRF2 could also estimate the effects of the mandatory clinical variables as well as the Cox model (Supplementary Figure \ref{figS:simLow-xi-misspecification}). 
When comparing the performances of the three GPTCMs which only model the cell-type-specific survival (GPTCM-noBVS1, GPTCM-Ber1 and GPTCM-MRF1) and the three GPTCMs that model both, the cell-type-specific survival and proportions in both low and high dimensions (GPTCM-noBVS2, GPTCM-Ber2 and GPTCM-MRF2), see Figures \ref{fig:simLow-brier} and \ref{fig:simHigh-brier}, it was interesting to observe that additionally modeling the proportions data can help for long-term survival prognosis, i.e., GPTCM-noBVS2, GPTCM-Ber2 and GPTCM-MRF2 outperform GPTCM-noBVS1, GPTCM-Ber1 and GPTCM-MRF1. 
This implies that cancer cell-type proportions (together with cell type-specific omics data) are useful for predicting cancer patient survival. 
Note that if only using bulk sequencing omics data are available, without single-cell RNA-seq or other data that allows distinguishing between different cell types, it is difficult to identify cell type-specific genes that are useful for guiding personalized treatment strategies to improve cancer patient survival. 

In the simulations with high-dimensional covariates, both GPTCM-Ber2 and GPTCM-MRF2 identified truly related and truly unrelated covariates linked to the cell-type-specific proportions. 
We attribute the very good performance of these models to the fact that the cell proportions are compositional data, and thus the Dirichlet model is an appropriate choice as a probabilistic model for such data. 
We have also shown that the GPTCM with an MRF prior can improve the identification of truly related cell-type-specific covariates if there is good prior knowledge for the biological network between omics variables. 

There are several other possibilities for further work that could improve the proposed models. 
For example, all the proposed Bayesian GPTCMs are parametric models. 
An alternative to the fully parametric settings would be to use a semiparametric functional form \citep{Hermansen2025} 
instead of the parametric Weibull survival function for $S_l(t)$ in GPTCM (\ref{eq:gptcm}), which may improve the survival prediction under model misspecification as shown in Supplementary Figure \ref{figS:simLow-brier-misspecification} while retaining good variable selection performance. 

Note that both the promotion time cure model (PTCM) and GPTCM are motivated by analyzing the mathematical dynamics of clonogenic tumor cells. 
However, the survival time or tumor growth should be also related to other types of cells, for example, immune cells. 
A more general and realistic model should account for the entire tumor microenvironment, for example, elucidating intercellular communication by modeling the interactions between cancer cell subtypes and immune cell subtypes and their omics information based on the single-cell and spatially resolved data of cancer tissues \citep{Longo2021,Gulati2024,Shang2025}. 
Also, the GPTCMs assume that the cell types are well categorized. 
But the discrete categorization based on single-cell RNA sequencing data tends to be an unsatisfactory model of the underlying biology \citep{AhlmannEltze2025}, a latent embedding for cell-type categorization jointly with GPTCM can be for future improvements. 
Another limitation of the GPTCM modeling framework is that cell type proportions data are only from the diagnosis time of cancer patients. 
It would be ideal to obtain longitudinal information of the proportions data, which may more accurately model the tumor evolution and predict patient survival \citep{Spitzer2025}.

\section*{Supplementary material}

The supplementary materials contain the Bayesian posterior inference with full posterior conditional distributions and Metropolis-Hastings algorithms, and include additional simulation results. 
The R package {\bf GPTCM} that implements all the proposed Bayesian GPTCMs is available on CRAN \url{https://CRAN.R-project.org/package=GPTCM}.

\section*{Acknowledgments}

This work was supported by the ERA PerMed under the ERA-NET Cofund scheme of the European Union's Horizon 2020 research and innovation framework program (grant ‘SYMMETRY’ ERAPERMED2021-330). 

\bibliographystyle{apalike}
\bibliography{refs}

\begin{thebibliography}{}

\bibitem[Ahlmann-Eltze and Huber, 2025]{AhlmannEltze2025}
Ahlmann-Eltze, C. and Huber, W. (2025).
\newblock Analysis of multi-condition single-cell data with latent embedding
  multivariate regression.
\newblock {\em Nature Genetics}, 57(3):659--667.

\bibitem[Bender et~al., 2005]{Bender2005}
Bender, R., Augustin, T., and Blettner, M. (2005).
\newblock Generating survival times to simulate {Cox} proportional hazards
  models.
\newblock {\em Statistics in Medicine}, 24(11):1713--1723.

\bibitem[Cai et~al., 2024]{Cai2024}
Cai, Y., Lu, Z., Chen, C., Zhu, Y., Chen, Z., Wu, Z., Peng, J., Zhu, X., Liu,
  Z., Li, B., Zhang, M., Huang, J., Li, Y., Liu, Y., Ma, Q., He, C., Chen, S.,
  Tian, W., Fan, L., Ning, C., Geng, H., Xu, B., Li, H., Zhu, X., Fang, J.,
  Wang, X., Zhang, S., Jin, M., Huang, C., Yang, X., Tian, J., and Miao, X.
  (2024).
\newblock An atlas of genetic effects on cellular composition of the tumor
  microenvironment.
\newblock {\em Nature Immunology}, 25(10):1959--1975.

\bibitem[Chen et~al., 1999]{Chen1999}
Chen, M.-H., Ibrahim, J.~G., and Sinha, D. (1999).
\newblock A new {B}ayesian model for survival data with a surviving fraction.
\newblock {\em Journal of the American Statistical Association},
  94(447):909--919.

\bibitem[Cooner et~al., 2007]{Cooner2007}
Cooner, F., Banerjee, S., Carlin, B.~P., and Sinha, D. (2007).
\newblock Flexible cure rate modeling under latent activation schemes.
\newblock {\em Journal of the American Statistical Association},
  102(478):560--572.

\bibitem[Cox, 1972]{Cox1972}
Cox, D.~R. (1972).
\newblock Regression models and life-tables.
\newblock {\em Journal of the Royal Statistical Society Series B: Statistical
  Methodology}, 34(2):187--202.

\bibitem[Dagogo-Jack and Shaw, 2017]{DagogoJack2017}
Dagogo-Jack, I. and Shaw, A.~T. (2017).
\newblock Tumour heterogeneity and resistance to cancer therapies.
\newblock {\em Nature Reviews Clinical Oncology}, 15(2):81–94.

\bibitem[Faucett and Thomas, 1996]{Faucett1996}
Faucett, C.~L. and Thomas, D.~C. (1996).
\newblock Simultaneously modelling censored survival data and repeatedly
  measured covariates: A {G}ibbs sampling approach.
\newblock {\em Statistics in Medicine}, 15(15):1663--1685.

\bibitem[Gilks et~al., 1995]{Gilks1995}
Gilks, W.~R., Best, N.~G., and Tan, K. K.~C. (1995).
\newblock Adaptive rejection {M}etropolis sampling within {G}ibbs sampling.
\newblock {\em Journal of the Royal Statistical Society Series C: Applied
  Statistics}, 44(4):455--472.

\bibitem[Gulati et~al., 2024]{Gulati2024}
Gulati, G.~S., D’Silva, J.~P., Liu, Y., Wang, L., and Newman, A.~M. (2024).
\newblock Profiling cell identity and tissue architecture with single-cell and
  spatial transcriptomics.
\newblock {\em Nature Reviews Molecular Cell Biology}, 26(1):11--31.

\bibitem[Gómez et~al., 2023]{Gomez2023}
Gómez, Y.~M., Gallardo, D.~I., Bourguignon, M., Bertolli, E., and Calsavara,
  V.~F. (2023).
\newblock A general class of promotion time cure rate models with a new
  biological interpretation.
\newblock {\em Lifetime Data Analysis}, 29(1):66--86.

\bibitem[Hermansen et~al., 2025]{Hermansen2025}
Hermansen, T.~O., Zucknick, M., and Zhao, Z. (2025).
\newblock Bayesian {C}ox model with graph-structured variable selection priors
  for multi-omics biomarker identification.
\newblock arXiv:2503.13078.

\bibitem[Kim et~al., 2011]{Kim2011}
Kim, S., Chen, M.-H., and Dey, D.~K. (2011).
\newblock A new threshold regression model for survival data with a cure
  fraction.
\newblock {\em Lifetime Data Analysis}, 17(1):101--122.

\bibitem[Li and Zhang, 2010]{Li2010}
Li, F. and Zhang, N.~R. (2010).
\newblock Bayesian variable selection in structured high-dimensional covariate
  spaces with applications in genomics.
\newblock {\em Journal of the American Statistical Association},
  105(491):1202--1214.

\bibitem[Longo et~al., 2021]{Longo2021}
Longo, S.~K., Guo, M.~G., Ji, A.~L., and Khavari, P.~A. (2021).
\newblock Integrating single-cell and spatial transcriptomics to elucidate
  intercellular tissue dynamics.
\newblock {\em Nature Reviews Genetics}, 22(10):627--644.

\bibitem[Ma and Yin, 2008]{Ma2008}
Ma, Y. and Yin, G. (2008).
\newblock Cure rate model with mismeasured covariates under transformation.
\newblock {\em Journal of the American Statistical Association},
  103(482):743--756.

\bibitem[Maier, 2014]{Maier2014}
Maier, M.~J. (2014).
\newblock {DirichletReg: D}irichlet regression for compositional data in {R}.
\newblock {Research Report Series{\slash}Department of Statistics and
  Mathematics} 125, {WU Vienna University of Economics and Business}, {Vienna}.

\bibitem[McGranahan and Swanton, 2017]{McGranahan2017}
McGranahan, N. and Swanton, C. (2017).
\newblock Clonal heterogeneity and tumor evolution: Past, present, and the
  future.
\newblock {\em Cell}, 168(4):613–628.

\bibitem[Neal, 2003]{Neal2003}
Neal, R.~M. (2003).
\newblock Slice sampling.
\newblock {\em The Annals of Statistics}, 31(3):705--767.

\bibitem[Shang et~al., 2025]{Shang2025}
Shang, L., Wu, P., and Zhou, X. (2025).
\newblock Statistical identification of cell type-specific spatially variable
  genes in spatial transcriptomics.
\newblock {\em Nature Communications}, 16(1).

\bibitem[Simon et~al., 2011]{Simon2011}
Simon, N., Friedman, J., Hastie, T., and Tibshirani, R. (2011).
\newblock Regularization paths for {Cox}’s proportional hazards model via
  coordinate descent.
\newblock {\em Journal of Statistical Software}, 39(5):1--13.

\bibitem[Spitzer et~al., 2025]{Spitzer2025}
Spitzer, A., Johnson, K.~C., Nomura, M., Garofano, L., Nehar-belaid, D.,
  Darnell, N.~G., Greenwald, A.~C., Bussema, L., Oh, Y.~T., Varn, F.~S.,
  D’Angelo, F., Gritsch, S., Anderson, K.~J., Migliozzi, S., Gonzalez~Castro,
  L.~N., Chowdhury, T., Robine, N., Reeves, C., Park, J.~B., Lipsa, A., Hertel,
  F., Golebiewska, A., Niclou, S.~P., Nusrat, L., Kellet, S., Das, S., Moon,
  H.-E., Paek, S.~H., Bielle, F., Laurenge, A., Di~Stefano, A.~L., Mathon, B.,
  Picca, A., Sanson, M., Tanaka, S., Saito, N., Ashley, D.~M., Keir, S.~T.,
  Ligon, K.~L., Huse, J.~T., Yung, W. K.~A., Lasorella, A., Iavarone, A.,
  Verhaak, R. G.~W., Tirosh, I., and Suvà, M.~L. (2025).
\newblock Deciphering the longitudinal trajectories of glioblastoma ecosystems
  by integrative single-cell genomics.
\newblock {\em Nature Genetics}, 57(5):1168--1178.

\bibitem[Stingo et~al., 2011]{Stingo2011}
Stingo, F.~C., Chen, Y.~A., Tadesse, M.~G., and Vannucci, M. (2011).
\newblock Incorporating biological information into linear models: {A B}ayesian
  approach to the selection of pathways and genes.
\newblock {\em The Annals of Applied Statistics}, 5(3):1978--2002.

\bibitem[Sutton and Barto, 1998]{Sutton1998}
Sutton, R. and Barto, A. (1998).
\newblock {\em {Reinforcement Learning: An Introduction}}.
\newblock MIT Press.

\bibitem[Thompson, 1933]{Thompson1933}
Thompson, W.~R. (1933).
\newblock On the likelihood that one unknown probability exceeds another in
  view of the evidence of two samples.
\newblock {\em Biometrika}, 25(3--4):285--294.

\bibitem[Vehtari et~al., 2024]{Vehtari2024}
Vehtari, A., Simpson, D., Gelman, A., Yao, Y., and Gabry, J. (2024).
\newblock Pareto smoothed importance sampling.
\newblock {\em Journal of Machine Learning Research}, 25:1--58.

\bibitem[Yakovlev et~al., 1996]{Yakovlev1996}
Yakovlev, A.~Y., Tsodikov, A.~D., and Asselain, B. (1996).
\newblock {\em Stochastic Models of Tumor Latency and Their Biostatistical
  Applications}.
\newblock World Scientific, Singapore.

\bibitem[Zhang et~al., 2022]{Zhang2022}
Zhang, A., Miao, K., Sun, H., and Deng, C.-X. (2022).
\newblock Tumor heterogeneity reshapes the tumor microenvironment to influence
  drug resistance.
\newblock {\em International Journal of Biological Sciences}, 18(7):3019--3033.

\bibitem[Zhao, 2025]{Zhao2025cran}
Zhao, Z. (2025).
\newblock {\em {GPTCM: Generalized Promotion Time Cure Model with Bayesian
  Shrinkage Priors}}.
\newblock R package version 1.1.2.

\bibitem[Zhao et~al., 2024a]{Zhao2024mrf}
Zhao, Z., Banterle, M., Lewin, A., and Zucknick, M. (2024a).
\newblock Multivariate {B}ayesian structured variable selection for
  pharmacogenomic studies.
\newblock {\em Journal of the Royal Statistical Society Series C: Applied
  Statistics}, 73(2):420--443.

\bibitem[Zhao and Kızılaslan, 2024]{Zhao2024gptcm}
Zhao, Z. and Kızılaslan, F. (2024).
\newblock A note on promotion time cure models with a new biological
  consideration.
\newblock arXiv:2408.17188.

\bibitem[Zhao et~al., 2024b]{Zhao2024survomics}
Zhao, Z., Zobolas, J., Zucknick, M., and Aittokallio, T. (2024b).
\newblock Tutorial on survival modeling with applications to omics data.
\newblock {\em Bioinformatics}, 40(3):btae132.

\end{thebibliography}

\clearpage
\setcounter{page}{1}
\pagenumbering{roman}
\renewcommand\thefigure{S\arabic{figure}}    
\setcounter{figure}{0}  
\renewcommand\thetable{S\arabic{table}}    
\setcounter{table}{0}  
\renewcommand\thesection{S\arabic{section}}    
\setcounter{section}{0}  

\title{\LARGE\centering Supplementary materials for \\
``Generalized promotion time cure model: A new modeling framework to identify cell-type-specific genes and improve survival prognosis''}

\bigskip
\section{Posterior inference}\label{secSuppl:posterior}

The joint posterior distribution is composed by the likelihood and the joint prior, see Section \ref{sec:computation}. 
Based on that, all available full posterior conditional distributions are as follows. 
\begin{flalign*}
f(\beta_{jl}| \gamma_{jl} = 1, \ \textbf{---}) 
&\propto \prod_{i=1}^n \Biggl\{
  e^{-\theta_i(1-\sum_{l'} \mathtt p_{il'} e^{- (t_i/\lambda_{il'})^\kappa})} 
  \theta_i\kappa t_i^{\kappa-1}\sum_{l'} \mathtt p_{il'}\lambda_{il'}^{-\kappa} e^{-(t_i/\lambda_{il'})^\kappa}
\Biggl\}^{\delta_i} &&\\
&\hspace{10mm} \times \Biggl\{ 
  e^{-\theta_i(1-\sum_{l'} \mathtt p_{il'} e^{-(t_i/\lambda_{il'})^\kappa})} 
  \Biggl\}^{1-\delta_i} 
\times e^{-\frac{\beta_{jl}^2}{2\tau_l^2}} &&\\
& \propto \prod_{i=1}^n \Biggl\{
  \sum_{l'} \frac{\kappa}{\lambda_{il'}} \left(\frac{t_i}{\lambda_{il'}}\right)^{\kappa-1} \mathtt p_{il'}e^{-(t_i/\lambda_{il'})^\kappa}
\Biggl\}^{\delta_i}
  e^{\theta_i \mathtt p_{il} e^{-(t_i/\lambda_{il})^\kappa}
  -\frac{\beta_{jl}^2}{2\tau_l^2}}, \ j=1,...,p,\ l=1,...,L, &&
\end{flalign*}
where $\log\lambda_{il} = \frac{\mu_{il}}{\Gamma(1+1/\kappa)}$, $\mu_{il} = \beta_{0l}+\mathbf X_{il}\bm\beta_l$. 
Note that the conditional part with notation ``\textbf{---}'' means all relevant parameters. 
If $\gamma_{jl} = 0$, then $\beta_{jl}=0$ without the need of a posterior conditional distribution. 
\begin{flalign*}
f(\beta_{0l}|\ \textbf{---}) 
&\propto \prod_{i=1}^n \Biggl\{
  e^{-\theta_i(1-\sum_{l'} \mathtt p_{il'} e^{- (t_i/\lambda_{il'})^\kappa})} 
  \theta_i\kappa t_i^{\kappa-1}\sum_{l'} \mathtt p_{il'}\lambda_{il'}^{-\kappa} e^{-(t_i/\lambda_{il'})^\kappa}
\Biggl\}^{\delta_i} &&\\
&\hspace{10mm} \times \Biggl\{ 
  e^{-\theta_i(1-\sum_{l'} \mathtt p_{il'} e^{-(t_i/\lambda_{il'})^\kappa})} 
  \Biggl\}^{1-\delta_i} 
\times e^{-\frac{\beta_{0l}^2}{2\bm\tau_0^2}} &&\\
& \propto \prod_{i=1}^n \Biggl\{
  \sum_{l'} \frac{\kappa}{\lambda_{il'}} \left(\frac{t_i}{\lambda_{il'}}\right)^{\kappa-1} \mathtt p_{il'}e^{-(t_i/\lambda_{il'})^\kappa}
\Biggl\}^{\delta_i}
  e^{\theta_i \mathtt p_{il} e^{-(t_i/\lambda_{il})^\kappa}
  -\frac{\beta_{0l}^2}{2\tau_0^2}}, \ j=1,...,p,\ l=1,...,L. &&
\end{flalign*}

\begin{flalign*}
f(\pi_{jl}| \ \textbf{---}) \propto \prod_{l'=1}^L \prod_{j'=1}^p f(\gamma_{j'l'} | \pi_{j'l'}) 
\cdot \mathcal Beta(a_\pi, b_\pi) 
\sim \mathcal Beta(a_\pi+\gamma_{jl}, b_\pi+p-\gamma_{jl}) &&
\end{flalign*}

\begin{flalign*}
  f(\tau_l^2| \ \textbf{---}) 
  &\propto \prod_{l'=1}^L \prod_{j=1}^p f(\beta_{jl'} | \gamma_{jl'}=1, \tau_{l'}^2) 
  \cdot \mathcal{IG}amma(a_\tau, b_\tau) 
  \sim \mathcal{IG}amma\left( a_\tau+0.5\sum_j\gamma_{jl},\ b_\tau+0.5\sum_j\beta_{jl}^2 \right) &&
\end{flalign*}

\begin{flalign*}
f(\xi_j|  \ \textbf{---}) 
&\propto 
\prod_{i=1}^n \Biggl\{
  e^{-\theta_i(1-\sum_l \mathtt p_{il} S_l(t_i))} 
  \theta_i\kappa t_i^{\kappa-1}\sum_l \mathtt p_{il}\lambda_{il}^{-\kappa} S_l(t_i)
\Biggl\}^{\delta_i}
\Biggl\{ 
  e^{-\theta_i(1-\sum_l \mathtt p_{il} S_l(t_i))} 
  \Biggl\}^{1-\delta_i} 
 e^{-\frac{\xi_j^2}{2v_j^2}} &&\\
 &\propto
 \prod_{i=1}^n \theta_i^{\delta_i}
   e^{-\theta_i(1-\sum_l \mathtt p_{il} S_l(t_i)) 
  -\frac{\xi_j^2}{2v_j^2}}, \ 
  \text{where } \theta_i = \exp\{\xi_0 + \mathbf X_{0i}\bm\xi\},\ \bm\xi = [\xi_1,...,\xi_d]^\top,\ j=1,...,d &&
\end{flalign*}

\begin{flalign*}
f(\xi_0|  \ \textbf{---}) \propto 
\prod_{i=1}^n \theta_i^{\delta_i}
e^{-\theta_i(1-\sum_l \mathtt p_{il} S_l(t_i))
-\frac{\xi_0^2}{2v_0^2}}, \ 
  \text{where } \theta_i = \exp\{\xi_0 + \mathbf X_{0i}\bm\xi\} &&
\end{flalign*}

\begin{flalign*}
  f(v_j^2| \ \textbf{---}) 
  \propto \prod_{j'=1}^d f(\xi_{j'} | v_{j'}^2) \cdot \mathcal{IG}amma( a_v,\ b_w) 
  \sim \mathcal{IG}amma\left( a_v+1,\ b_w+0.5\sum_{j'}\xi_{j'}^2 \right) &&
\end{flalign*}

\begin{flalign*}
  f(\kappa| \ \textbf{---}) 
  &\propto  
  \prod_{i=1}^n \Biggl\{
    e^{-\theta_i(1-\sum_l \mathtt p_{il} S_l(t_i))} 
    \theta_i\kappa t_i^{\kappa-1}\sum_l \mathtt p_{il}\lambda_{il}^{-\kappa} S_l(t_i)
  \Biggl\}^{\delta_i}
  \Biggl\{ 
    e^{-\theta_i(1-\sum_l \mathtt p_{il} S_l(t_i))} 
    \Biggl\}^{1-\delta_i} 
  \cdot \mathcal{G}amma(a_\kappa, b_\kappa) &&\\
  & \propto  
  \prod_{i=1}^n \Biggl\{
    \sum_l \frac{\kappa}{\lambda_{il}} \left(\frac{t_i}{\lambda_{il}}\right)^{\kappa-1} \mathtt p_{il} S_l(t_i)
  \Biggl\}^{\delta_i}
    e^{\theta_i\sum_l \mathtt p_{il} S_l(t_i)} 
  \cdot \mathcal{G}amma(a_\kappa, b_\kappa) &&
\end{flalign*}

\begin{flalign*}
f(\zeta_{jl}|\eta_{jl} = 1, \ \textbf{---}) 
&\propto 
\prod_{i=1}^n \Biggl\{
  e^{-\theta_i(1-\sum_{l'} \mathtt p_{il'} S_{l'}(t_i))} 
  \theta_i\kappa t_i^{\kappa-1}\sum_{l'} \mathtt p_{il'}\lambda_{il'}^{-\kappa} S_{l'}(t_i)
\Biggl\}^{\delta_i}
\Biggl\{ 
  e^{-\theta_i(1-\sum_{l'} \mathtt p_{il'} S_{l'}(t_i))} 
  \Biggl\}^{1-\delta_i} &&\\
 &\ \ \ \cdot f(\tilde{\bm p}_i | \mathbf p_i, \phi)  
 \prod_{l'=1}^L \prod_{j'=1}^p f(\zeta_{j'l'} | \eta_{j'l'}=1, \rho_{j'l'}) &&\\
&\propto 
\prod_{i=1}^n \Biggl\{
  \sum_{l'} \frac{\kappa}{\lambda_{il'}} \left(\frac{t_i}{\lambda_{il'}}\right)^{\kappa-1} \mathtt p_{il'} S_{l'}(t_i)
\Biggl\}^{\delta_i}
  e^{\theta_i\sum_{l'} \mathtt p_{il'} S_{l'}(t_i)}
  \cdot \frac{\Gamma(\alpha_{i0})}{\prod_{l'}\Gamma(\alpha_{il'})} \prod_{l'}\tilde{\mathtt p}_{il'}^{\alpha_{il'}-1}
  \cdot e^{-\frac{\zeta_{jl}^2}{2w_l^2}} &&\\
  &\propto 
  \prod_{i=1}^n \Biggl\{
    \sum_{l'} \lambda_{il'}^{-\kappa} \frac{\alpha_{il'}}{\alpha_{i0}} S_{l'}(t_i)
  \Biggl\}^{\delta_i}
    e^{\theta_i\sum_{l'} \frac{\alpha_{il'}}{\alpha_{i0}} S_{l'}(t_i)}
    \cdot \frac{\Gamma(\alpha_{i0})}{\prod_{l'}\Gamma(\alpha_{il'})} \prod_{l'}\tilde{\mathtt p}_{il'}^{\alpha_{il'}-1}
    \cdot e^{-\frac{\zeta_{jl}^2}{2w_l^2}} &&
\end{flalign*}

\begin{flalign*}
  f(\zeta_{0l}| \ \textbf{---}) 
  &\propto 
\prod_{i=1}^n \Biggl\{
  \sum_{l'} \lambda_{il'}^{-\kappa}\frac{\alpha_{il'}}{\alpha_{i0}} \mathtt p_{il'} S_{l'}(t_i)
\Biggl\}^{\delta_i}
  e^{\theta_i\sum_{l'} \frac{\alpha_{il'}}{\alpha_{i0}} S_{l'}(t_i)}
  \cdot \frac{\Gamma(\alpha_{i0})}{\prod_{l'}\Gamma(\alpha_{il'})} \prod_{l'}\tilde{\mathtt p}_{il'}^{\alpha_{il'}-1}
  \cdot e^{-\frac{\zeta_{0l}^2}{2w_0^2}} &&
\end{flalign*}

\begin{flalign*}
f(\rho_{jl}|  \ \textbf{---}) \propto \prod_{l'=1}^L \prod_{j'=1}^p f(\eta_{j'l'} | \rho_{j'l'}) 
\cdot \mathcal Beta(a_\rho, b_\rho) 
\sim \mathcal Beta(a_\rho+\eta_{jl}, b_\rho+p-\eta_{jl}) &&
\end{flalign*}

\begin{flalign*}
  f(w_l^2| \ \textbf{---}) 
  \propto \prod_{l'=1}^L \prod_{j=1}^p f(\zeta_{jl'} | \eta_{jl'}=1, w_{l'}^2) 
  \cdot \mathcal{IG}amma(a_w, b_w) 
  \sim \mathcal{IG}amma\left( a_w+0.5\sum_{j}\eta_{jl},\ b_w+0.5\sum_{j}\zeta_{jl}^2 \right) &&
\end{flalign*}

\begin{flalign*}
  f(w_0^2|\ \textbf{---}) 
  \propto \prod_{l=1}^L f(\zeta_{0l} | w_0^2) 
  \cdot \mathcal{IG}amma(a_{w_0}, b_{w_0})
  \sim  \mathcal{IG}amma\left( a_{w_0}+0.5L,\ b_{w_0}+0.5\sum_{l}\zeta_{0l}^2 \right) &&
\end{flalign*}

\medskip
\noindent{\bf $\bullet$ $\gamma$ sampling}
\medskip

For a randomly selected $l$-th cell type, $l\in\{1,...,L\}$, we use the Metropolis-Hastings algorithm to update the pair $(\bm\beta_l,\bm\gamma_l)$ with acceptance probability
$$
\min \left\{ 1, \frac{f(\bm\gamma_l^*) f(\mathcal D | \bm\gamma_l^*,\bm\beta_l^*,\textbf{---})}
    {f(\bm\gamma_l) f(\mathcal D | \bm\gamma_l,\bm\beta_l,\textbf{---})} \times 
    \frac{q(\bm\gamma_l| \bm\gamma_l^*)}
    {q(\bm\gamma_l^*| \bm\gamma_l)}
    \right\},
$$
where $\bm\gamma_l^*$ is the proposal variable selection indicators, and $\bm\beta_l^*$ is the proposal coefficients conditional on $\bm\gamma_l^*$ by ARMS. 
We can use the MC3 sampling or multi-armed bandit technique for obtaining a proposal $\bm\gamma_l^*$. 
With the MC3 sampling, we use symmetric proposal distribution $q_{\bm\gamma_l}(\bm\gamma_l | \bm\gamma_l^*) = q_{\bm\gamma_l}(\bm\gamma_l^* | \bm\gamma_l) \sim \mathcal Bernoulli(0.5)$. 
With the $\epsilon$-greedy strategy for a multi-armed bandit problem, we will evaluate all previous Bayesian variable selection in the proposal ratio $q(\bm\gamma_l| \bm\gamma_l^*)/q(\bm\gamma_l^*| \bm\gamma_l)$. 

\medskip
\noindent{\bf $\bullet$ $\eta$ sampling}
\medskip

Using the same approach as $\gamma$ sampling.

\clearpage
\section{Additional simulation results in low dimensions}

\begin{figure}[!htbp]
  \centering
  \includegraphics[height=0.74 \textwidth]{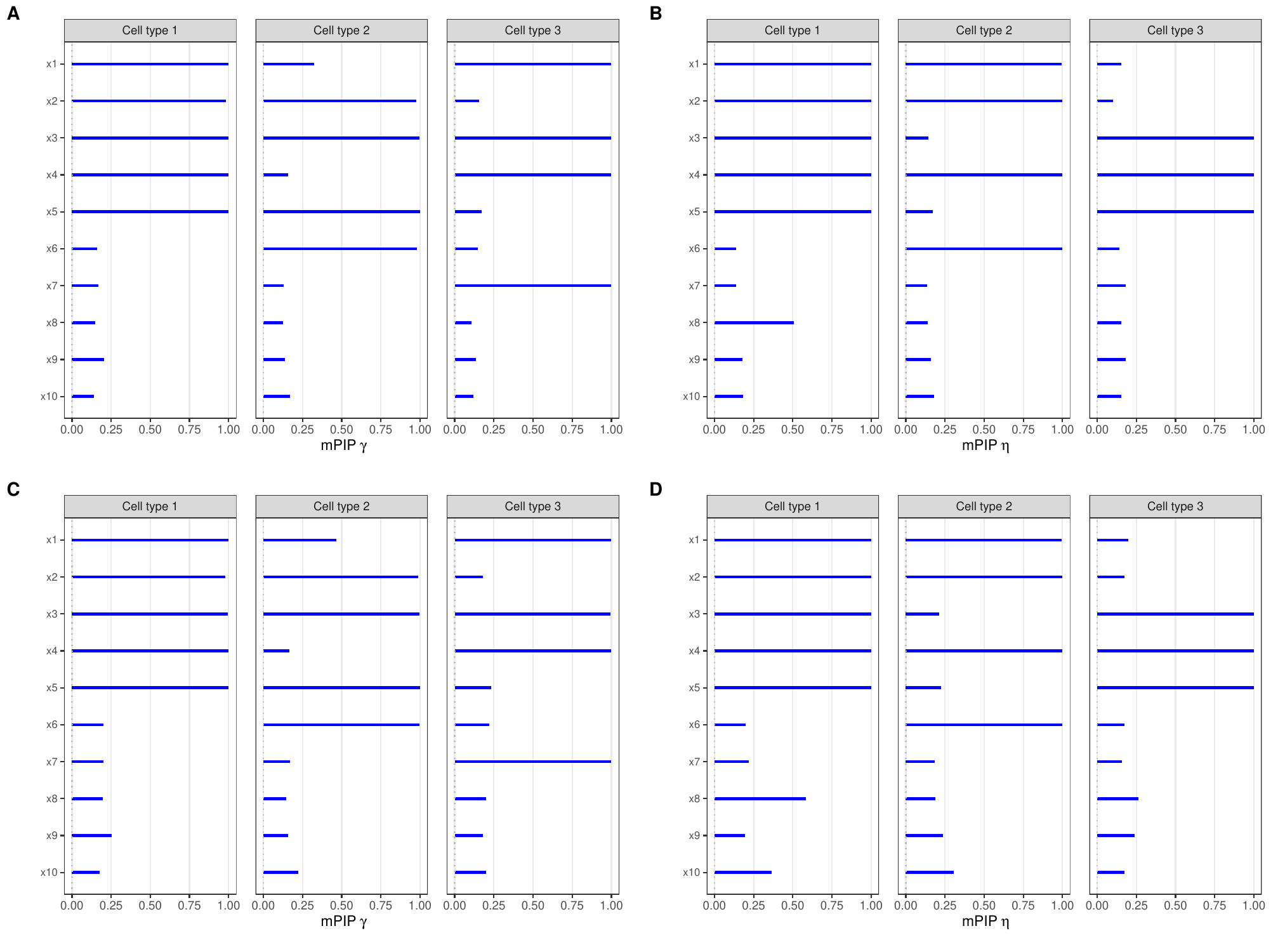} 
  \caption{\it Simulation results in low dimensions: Marginal posterior inclusion probabilities (mPIPs) of the variables linked to cell-type-specific survival and proportions by GPTCM-MRF2 and GPTCM-Ber2 based on one simulated data set. (A) mPIPs of the variables linked to cell-type-specific survival by GPTCM-MRF2. (B) mPIPs of the variables linked to cell-type-specific proportions by GPTCM-MRF2. (C) mPIPs of the variables linked to cell-type-specific survival by GPTCM-Ber2. (D) mPIPs of the variables linked to cell-type-specific proportions by GPTCM-Ber2. }
  \label{figS:simLow-mPIP}
\end{figure}

\begin{figure}
  \centering
  \makebox[\textwidth][c]{
  \includegraphics[height=0.4 \textwidth]{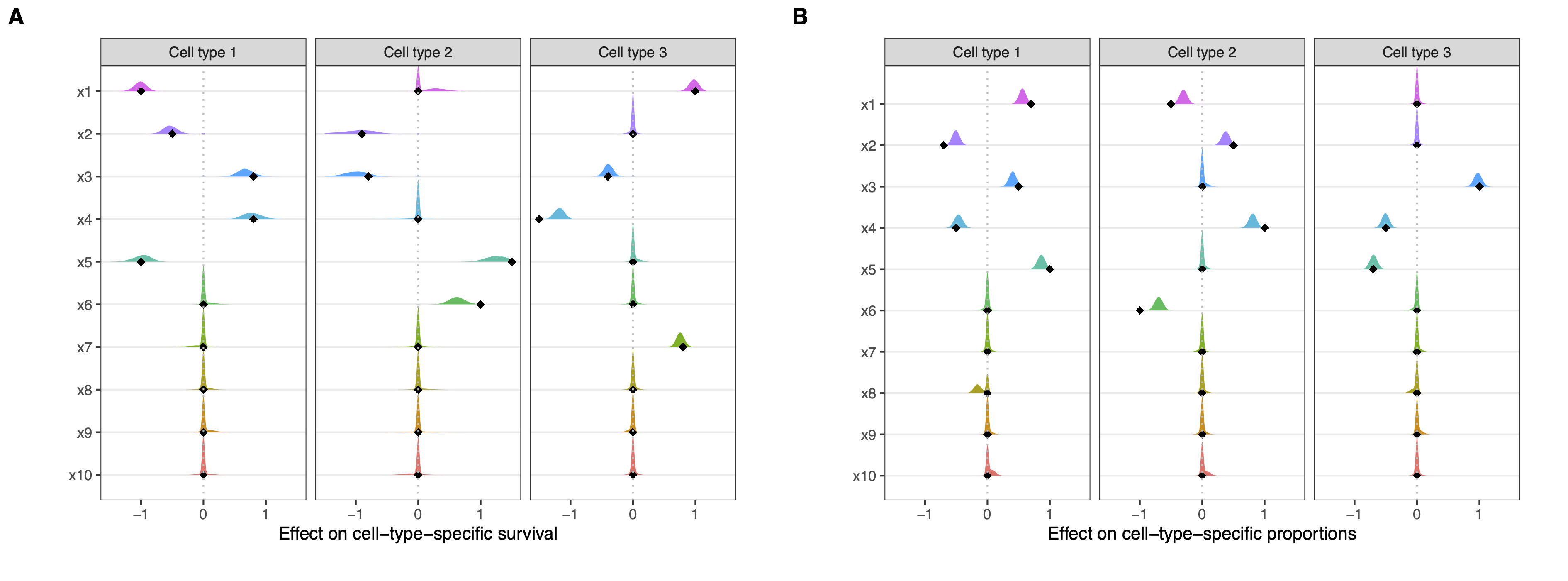} }
  \caption{\it Simulation results in low dimensions: Posterior distributions of the effects on cell-type-specific survival and proportions by GPTCM-Ber2 based on one simulated data set. The black colored diamond indicates the true effect. (A) Posterior distributions of the effects on cell-type-specific survival. (B) Posterior distributions of the effects on cell-type-specific proportions by the GPTCM-noBVS2.}
  \label{figS:simLow-postCoeff-gptcm4}
\end{figure}

\begin{figure}[!htbp]
  \centering
  \includegraphics[height=0.76 \textwidth]{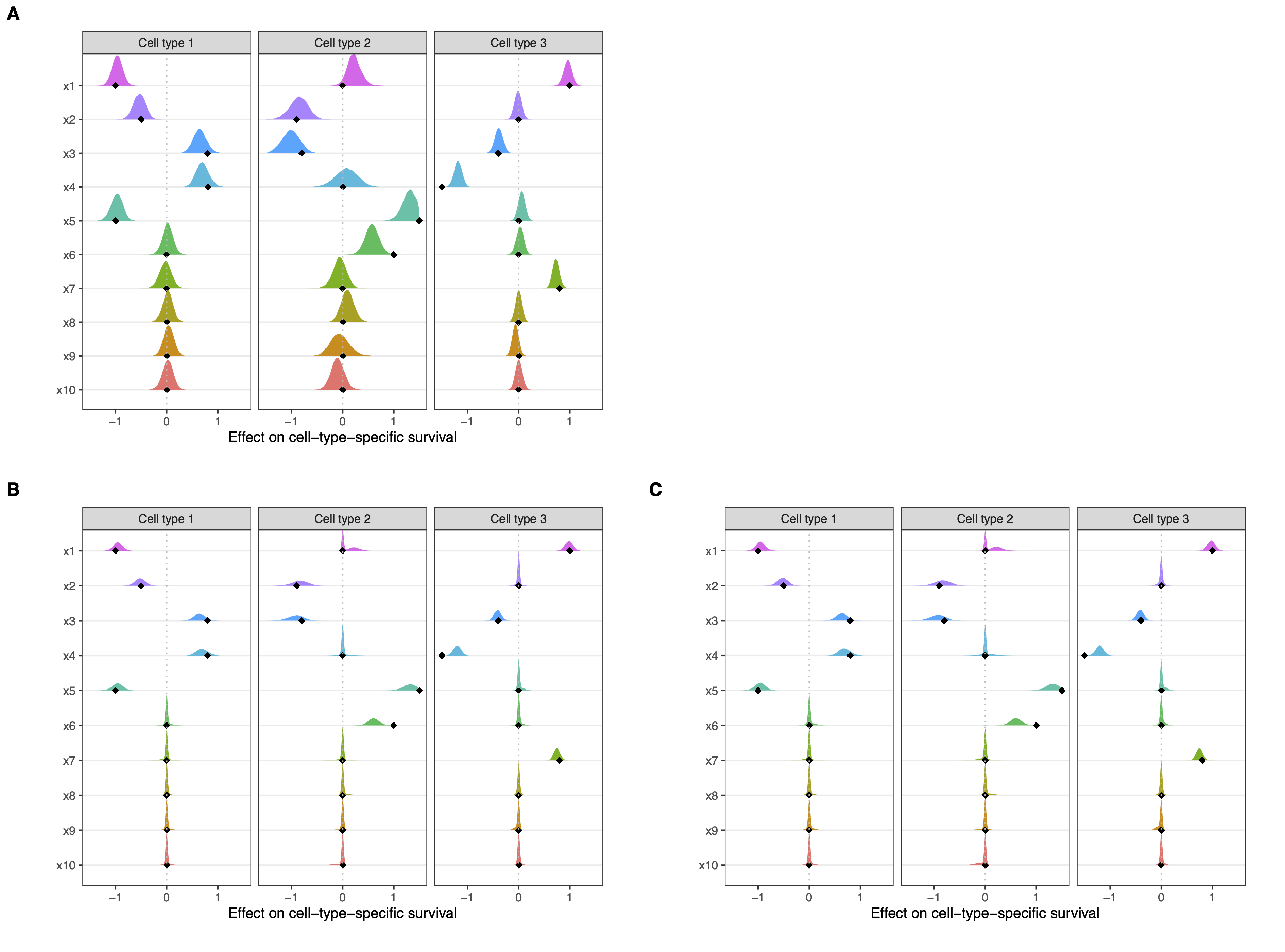} 
  \caption{\it Simulation results in low dimensions: Posterior distributions of the effects on cell-type-specific survival based on one simulated data set. The black colored diamond indicates the true effect. (A) Posterior distributions of the effects on cell-type-specific survival by GPTCM-noBVS1. (B) Posterior distributions of the effects on cell-type-specific survival by GPTCM-Ber1. (C) Posterior distributions of the effects on cell-type-specific proportions by GPTCM-MRF1.}
  \label{figS:simLow-postCoeff-gptcm135}
\end{figure}

\begin{figure}[!htbp]
  \centering
  \includegraphics[height=0.36 \textwidth]{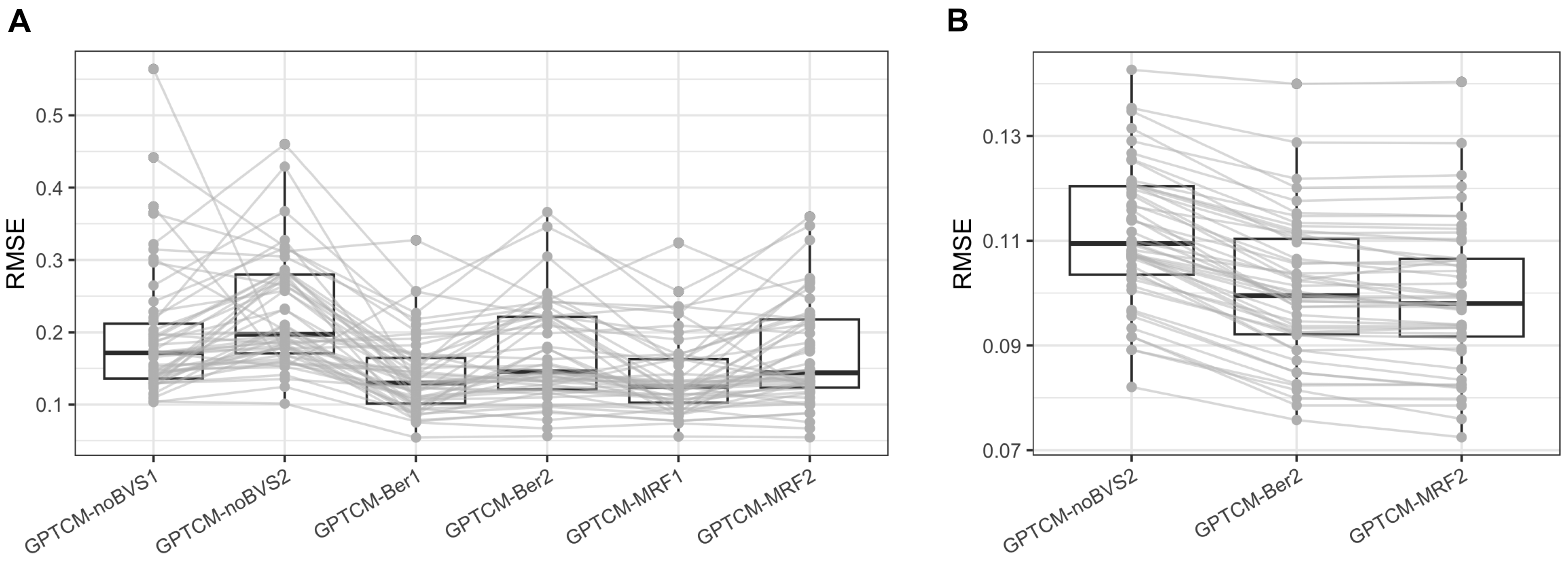} 
  \caption{\it Simulation results in low dimensions: 
  Root mean square errors (RMSE) of the estimated effects on cell-type-specific survival (i.e. $\frac{1}{\sqrt{pL}}\|\hat{\bm\beta}-\bm\beta\|_2$) and on cell-type-specific proportions (i.e. $\frac{1}{\sqrt{pL}}\|\hat{\bm\zeta}-\bm\zeta\|_2$) based on 50 simulated data sets. The gray lines connect points from the same simulated data. (A) RMSE of the estimated effects on cell-type-specific survival. (B) RMSE of the estimated effects on cell-type-specific proportions.}
  \label{figS:simLow-rmseBeta}
\end{figure}

\clearpage
\section{Simulation results in low dimensions for assessing model misspecification}

\begin{figure}[!htbp]
  \centering
  \includegraphics[height=0.5 \textwidth]{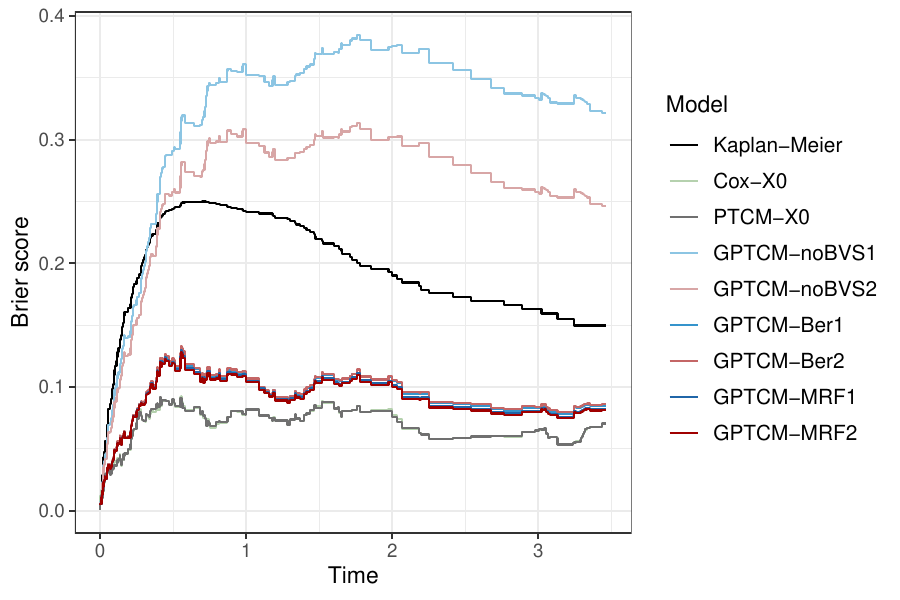}
  \caption{\it Simulation results in low dimensions under model misspecification: Prediction errors of classical survival models and GPTCMs in terms of time-dependent Brier score. The Kaplan-Meier method is as a reference that did not account for any covariate. The predictions of GPTCM-noBVS1 and GPTCM-noBVS2 were based on the posterior mean of all parameters. The predictions of other GPTCMs were based on the estimates of their median probability models (MPMs) for the effects of covariates with variable selection and estimates of posterior mean for other parameters. 
  }
  \label{figS:simLow-brier-misspecification}
\end{figure}

\begin{figure}[!htbp]
  \centering
  \makebox[\textwidth][c]{
  \includegraphics[height=0.75 \textwidth]{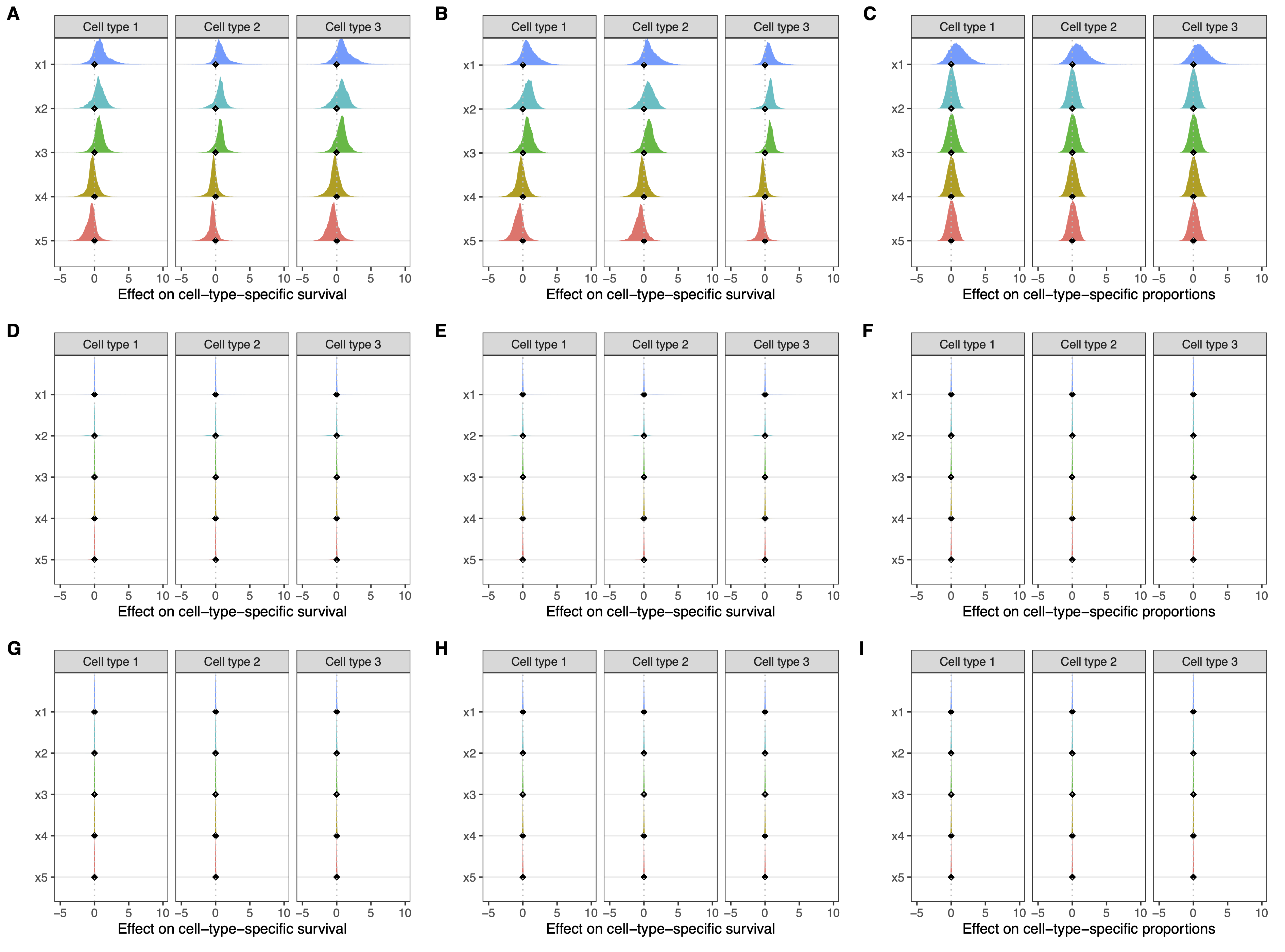} 
  }
  \caption{\it Simulation results in low dimensions under model misspecification: Posterior distributions of the effects on cell-type-specific survival and on cell-type-specific proportions. The black colored diamond indicates the true effect. (A) Posterior distributions of the effects on cell-type-specific survival by GPTCM-noBVS1. 
  (B) Posterior distributions of the effects on cell-type-specific survival by GPTCM-noBVS2. 
  (C) Posterior distributions of the effects on cell-type-specific proportions by GPTCM-noBVS2. 
  (D) Posterior distributions of the effects on cell-type-specific survival by GPTCM-Ber1. 
  (E) Posterior distributions of the effects on cell-type-specific survival by GPTCM-Ber2. 
  (F) Posterior distributions of the effects on cell-type-specific proportions by GPTCM-Ber2. 
  (G) Posterior distributions of the effects on cell-type-specific survival by GPTCM-MRF1. 
  (H) Posterior distributions of the effects on cell-type-specific survival by GPTCM-MRF2. 
  (I) Posterior distributions of the effects on cell-type-specific proportions by GPTCM-MRF2. 
  }
  \label{figS:simLow-postCoeff-misspecification}
\end{figure}

\begin{figure}[!htbp]
  \centering
  \includegraphics[height=0.6 \textwidth]{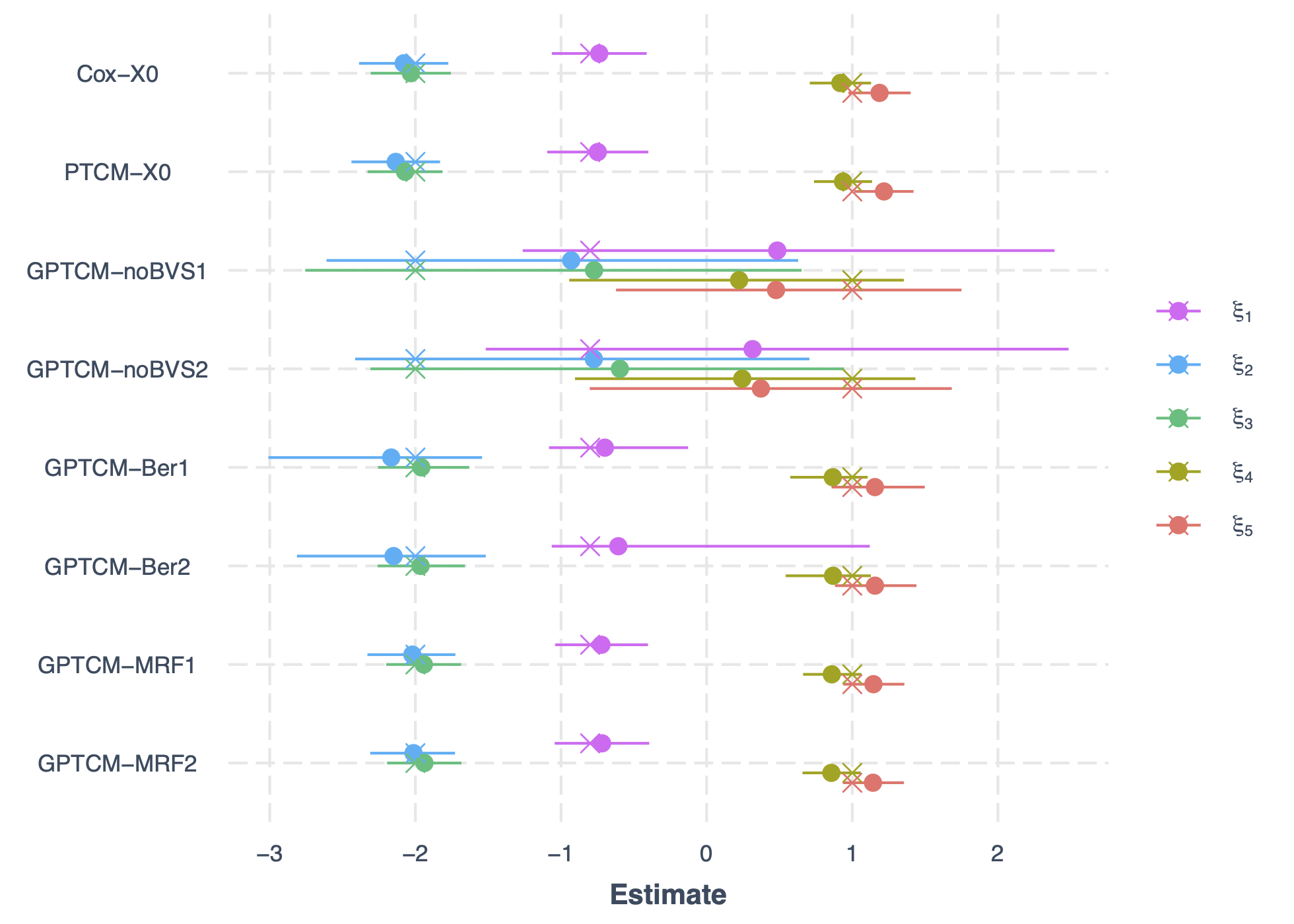} 
  \caption{\it Simulation results in low dimensions under model misspecification: Estimates of the clinical variables' effects on the cure fraction. The circle point shows a point estimate (posterior mean if it corresponds to a Bayesian GPTCM). The error bar shows a 95\% confidence or credible interval. The cross shape denotes the true effect of a clinical variable.}
  \label{figS:simLow-xi-misspecification}
\end{figure}

\clearpage
\section{Additional simulation results in high dimensions}

\begin{figure}[!htbp]
  \centering
  \includegraphics[height=0.36 \textwidth]{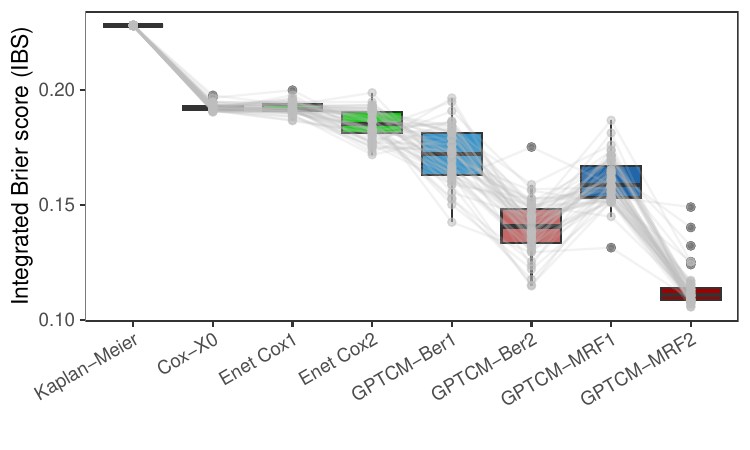}
  \caption{\it Simulation results in high dimensions: 
  Out-sample prediction errors of classical survival models and GPTCMs in terms of integrated Brier score (IBS) up to time point 1.2 trained by 50 simulated data sets. 
  The grey colored lines in panel A connect the performances of different models on the same data set. 
  The Kaplan-Meier method is as a reference that did not account for any covariate. 
  The elastic net Cox model ``Enet Cox1'' included the clinical variables $\mathbf X_0$ and mean aggregate variables of $\mathbf X_1,...,\mathbf X_L$.  The elastic net Cox model ``Enet Cox2'' included the clinical variables $\mathbf X_0$ and all cell-type-specific variables $[\mathbf X_1,...,\mathbf X_L]$. 
  The predictions of other GPTCMs were based on the estimates of their median probability models (MPMs) for the effects of covariates with variable selection and estimates of posterior mean for other parameters.}
  \label{figS:simHigh-brier1.2}
\end{figure}

\begin{figure}[!htbp]
  \centering
  \makebox[\textwidth][c]{
  \includegraphics[height=0.9 \textwidth]{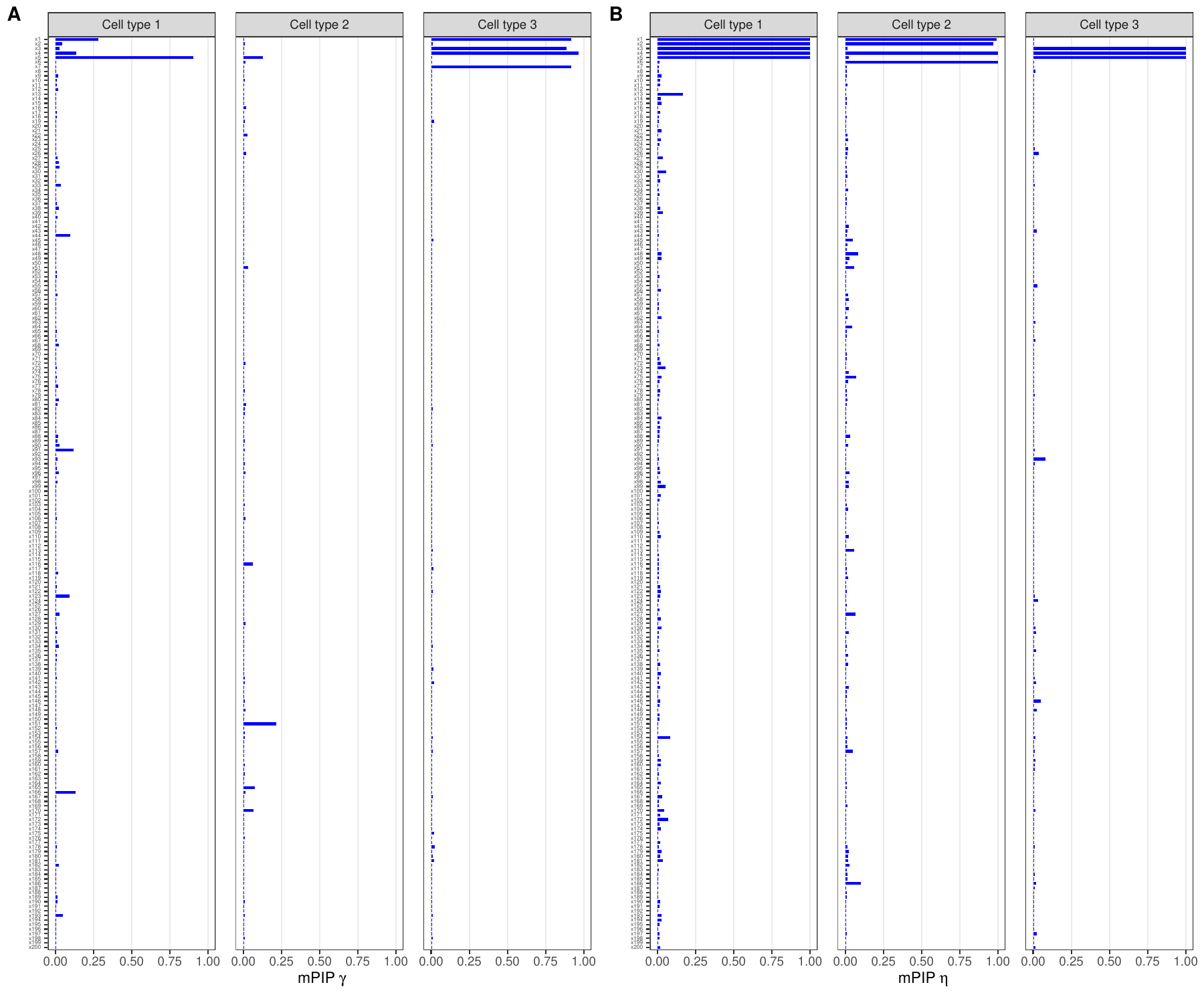} }
  \caption{\it Simulation results in high dimensions: Marginal posterior inclusion probabilities (mPIPs) of the variables linked to cell-type-specific survival and proportions by GPTCM-Ber2. (A) mPIPs of the variables linked to cell-type-specific survival. (B) mPIPs of the variables linked to cell-type-specific proportions.}
  \label{figS:simHigh-mPIP-gptcm4}
\end{figure}

\begin{figure}[!htbp]
  \centering
  \makebox[\textwidth][c]{
  \includegraphics[height=0.9 \textwidth]{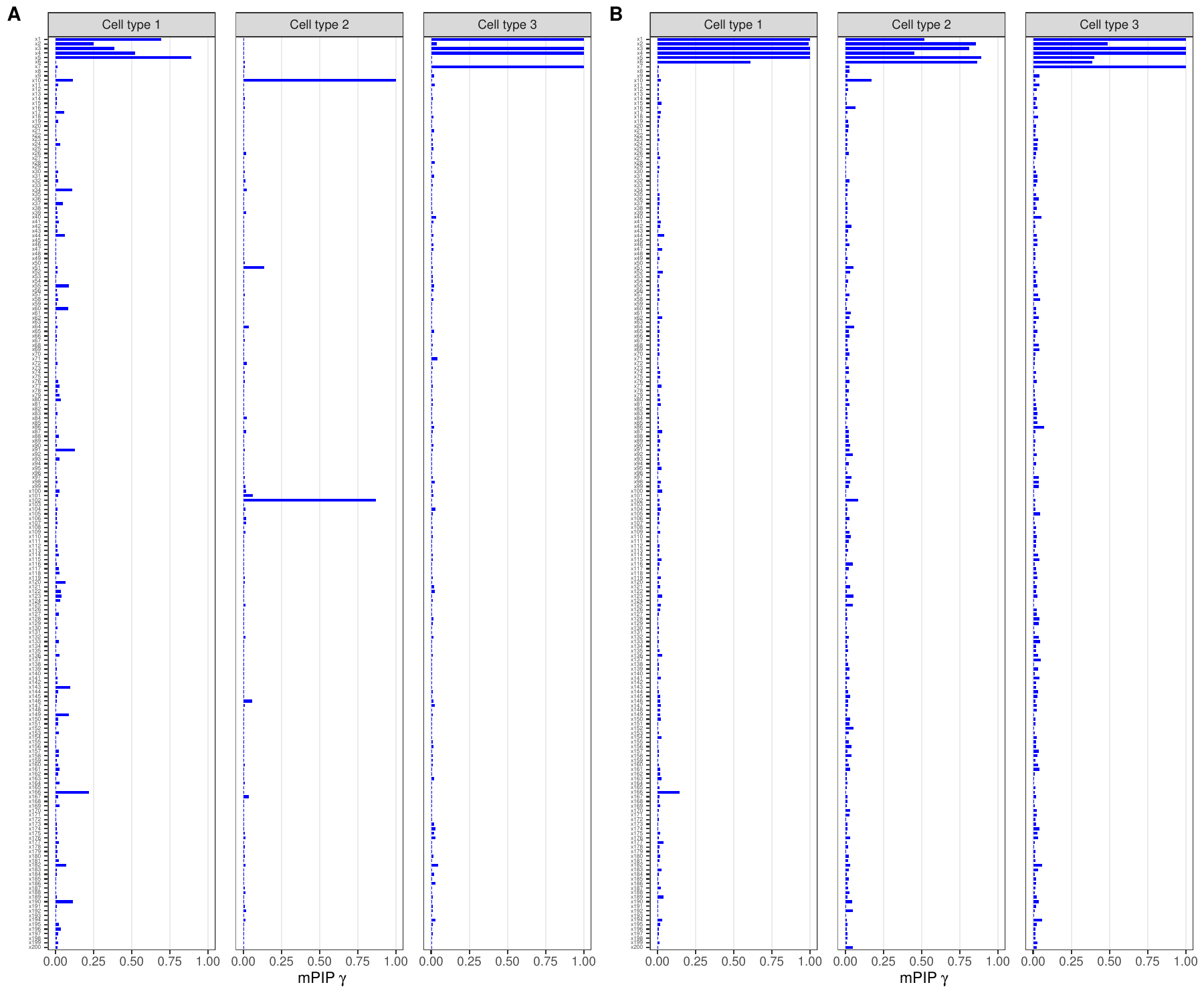} }
  \caption{\it Simulation results in high dimensions: Marginal posterior inclusion probabilities (mPIPs) of the variables linked to cell-type-specific survival and proportions by GPTCM-Ber1 and GPTCM-MRF1. (A) mPIPs of the variables linked to cell-type-specific survival by GPTCM-Ber1. (B) mPIPs of the variables linked to cell-type-specific survival by GPTCM-MRF1.}
  \label{figS:simHigh-mPIP-gptcm3_5}
\end{figure}

\begin{figure}[!htbp]
  \centering
  \makebox[\textwidth][c]{
  \includegraphics[height=0.6 \textwidth]{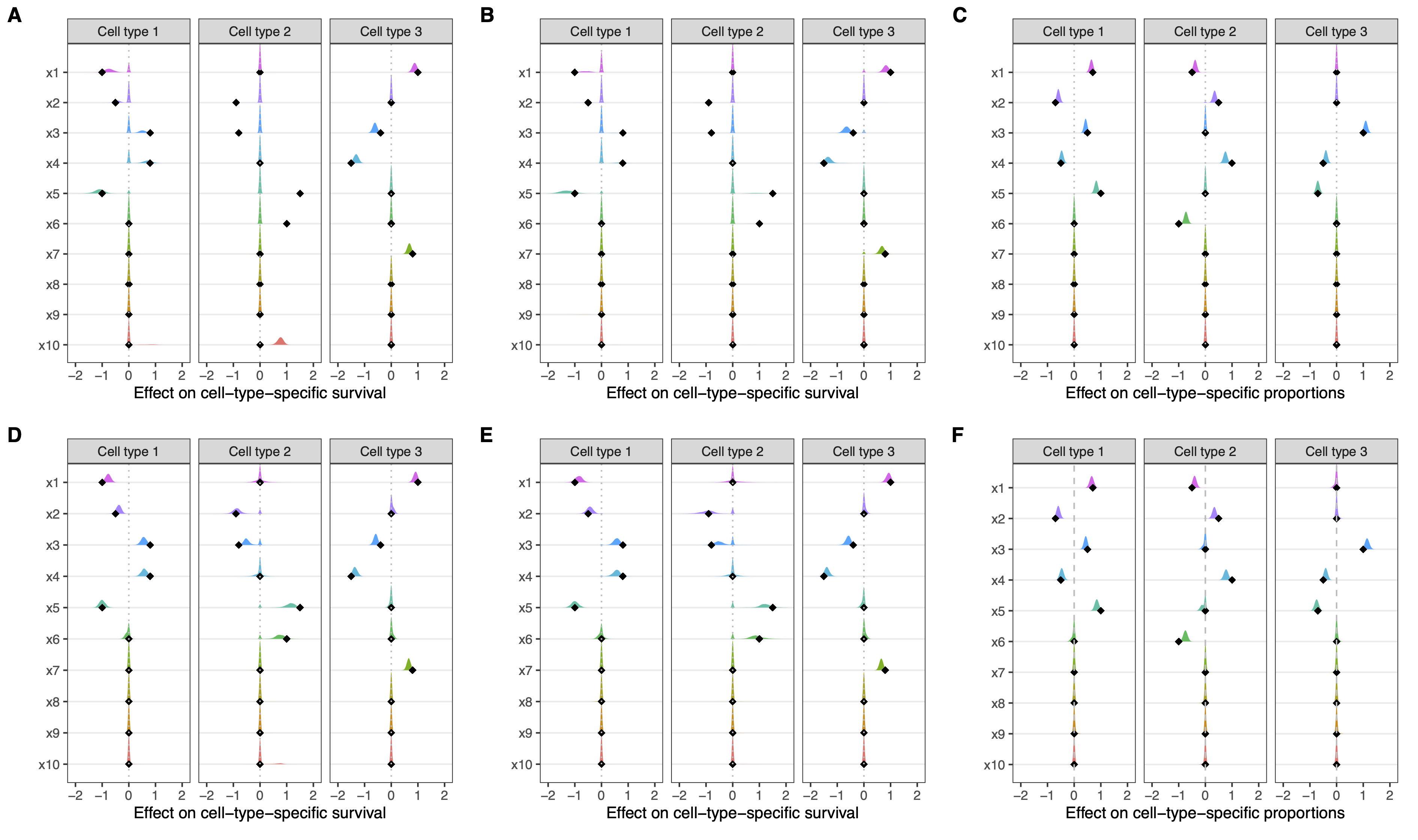} }
  \caption{\it Simulation results in high dimensions: The first 10 posterior distributions of the effects on cell-type-specific survival and proportions by GPTCM-Ber1, GPTCM-Ber2, GPTCM-MRF1, and GPTCM-MRF2. The black colored diamond indicates the true effect. 
  (A) Posterior distributions of the effects on cell-type-specific survival by GPTCM-Ber1. 
  (B) Posterior distributions of the effects on cell-type-specific survival by GPTCM-Ber2. 
  (C) Posterior distributions of the effects on cell-type-specific proportions by GPTCM-Ber2. 
  (D) Posterior distributions of the effects on cell-type-specific survival by GPTCM-MRF1. 
  (E) Posterior distributions of the effects on cell-type-specific survival by GPTCM-MRF2. 
  (F) Posterior distributions of the effects on cell-type-specific proportions by GPTCM-MRF2. }
  \label{figS:simHigh-postCoeff-gptcm135}
\end{figure}

\end{document}